\UseRawInputEncoding
\documentclass[12pt]{article}
\usepackage{amsmath,bm}
\usepackage{amssymb}
\usepackage{geometry}
\usepackage{graphicx}
\usepackage{hyperref}
\usepackage{subfig}
\usepackage{mciteplus}
\usepackage{float}
\usepackage{slashed}
\usepackage{multicol}

\usepackage[font=footnotesize]{caption}

\def\gtwid{\mathrel{\raise.3ex\hbox{$>$\kern-.75em\lower1ex\hbox{$\sim
$}}}}
\def\vio{\mathrel{\hbox{$E$\kern-.60em\hbox{$/
$}}}}
\newcommand{\newc}{\newcommand*}

\long\def\begincomment#1\endcomment{%
        \begingroup\sf\baselineskip12pt#1\endgroup}

\newc{\etal}{\textrm{et al.}} 
\newc{\eg}{\textrm{e.g.}} 
\newc{\ie}{\textrm{i.e.}}
\newc{\etc}{\textrm{etc}}
\newc\vs{\textrm{vs.}}
\newc{\cl}{\rm {C.L.}}

\newc{\ev}{\ensuremath{\,\mathrm{eV}}}
\newc{\kev}{\ensuremath{\,\mathrm{keV}}}
\newc{\mev}{\ensuremath{\,\mathrm{MeV}}}
\newc{\gev}{\ensuremath{\,\mathrm{GeV}}}
\newc{\tev}{\ensuremath{\,\mathrm{TeV}}}
\newc{\MeV}{\mev} 
\newc{\TeV}{\tev}
\newc{\invpb}{\ensuremath{/\text{pb}}}
\newc{\invfb}{\ensuremath{\,\textrm{fb}^{-1}}}
\newc\nb{\ensuremath{\,\mathrm{nb}}} \newc\pb{\ensuremath{\,\mathrm{pb}}} \newc\fb{\ensuremath{\,\mathrm{fb}}}
\newc\pc{\ensuremath{\,\mathrm{pc}}}
\newc\kpc{\ensuremath{\,\mathrm{kpc}}}
\newc\mpc{\ensuremath{\,\mathrm{Mpc}}}
\newc\ps{\ensuremath{\,\mathrm{ps}}} 
\newc\cmeter{\ensuremath{\,\mathrm{cm}}} 
\newc\meter{\ensuremath{\,\mathrm{m}}} 
\newc\kmeter{\ensuremath{\,\mathrm{km}}}
\newc\second{\ensuremath{\,\mathrm{s}}}
\newc\msecond{\ensuremath{\,\mathrm{ms}}}
\newc\nsecond{\ensuremath{\,\mathrm{ns}}}
\newc\psecond{\ensuremath{\,\mathrm{ps}}}

\newc{\chisqmin}{\ensuremath{\chi^2_{\mathrm{min}}}}
\newc{\Delchisq}{\ensuremath{\Delta\chi^2}}
\newc{\chisq}{\ensuremath{\chi^2}}
\newc{\like}{\ensuremath{\mathcal{L}}}

\newc\lsim{\ensuremath{\mathrel{\rlap{\lower4pt\hbox{\hskip1pt$\sim$}}\raise1pt\hbox{$<$}}}}
\newc\gsim{\ensuremath{\mathrel{\rlap{\lower4pt\hbox{\hskip1pt$\sim$}}\raise1pt\hbox{$>$}}}}
\newc{\VEV}[1]{\ensuremath{\langle #1 \rangle}}
\newc{\dl}{\ensuremath{\stackrel{\leftarrow}{D}}}
\newc{\dr}{\ensuremath{\stackrel{\rightarrow}{D}}}

\newc{\bcenter}{\begin{center}}    \newc{\ecenter}{\end{center}}
\newc{\bfl}{\begin{flushleft}}    \newc{\efl}{\end{flushleft}}
\newc{\bfr}{\begin{flushright}}    \newc{\efr}{\end{flushright}}

\newc{\bi}{\begin{itemize}}
\newc{\ei}{\end{itemize}}
\newc{\bed}{\begin{description}}
\newc{\eed}{\end{description}}
\newc{\ben}{\begin{enumerate}}
\newc{\een}{\end{enumerate}}

\newc{\be}{\begin{equation}}
\newc{\ee}{\end{equation}}
\newc{\bea}{\begin{eqnarray}}
\newc{\eea}{\end{eqnarray}}
\newc{\bfle}{\begin{flalign}}
\newc{\efle}{\end{flalign}}
\newc{\ra}{\rightarrow}

\newc{\alphas}{\ensuremath{\alpha_s}}
\newc{\alphatwo}{\ensuremath{\alpha_2}}
\newc{\alphaone}{\ensuremath{\alpha_1}}
\newc{\alphai}[1]{\ensuremath{\alpha_{#1}}}
\newc{\alphaem}{\ensuremath{\alpha_{\mathrm{em}}}}
\newc{\alphaeff}{\ensuremath{\alpha_{\mathrm{eff}}}}
\newc{\sineff}{\ensuremath{\sin \theta_{\mathrm{eff}}}}
\newc{\sinsqeff}{\ensuremath{\sin^2 \theta_{\mathrm{eff}}}}
\newc{\dalphahad}{\ensuremath{\Delta \alpha_{\mathrm{had}}}}
\newc{\yt}{\ensuremath{h_t}} \newc{\yb}{\ensuremath{h_b}} \newc{\ytau}{\ensuremath{h_{\tau}}}
\newc\mz{\ensuremath{M_Z}} 
\newc\mw{\ensuremath{m_W}}
\newc\mZ{\mz}        \newc\mW{\mw}
\newc\mhsm{\ensuremath{ m_{H_{\mathrm{SM}}}}}
\newc{\mtop}{\ensuremath{ m_t}}               \newc{\mtpole}{\ensuremath{ M_t}}
\newc{\mbottom}{\ensuremath{ m_b}} 
\newc{\mtau}{\ensuremath{ m_{\tau}}}
\newc{\mt}{\mtpole}
\newc{\mb}{\mbottom} 
\newc{\rtwogg}{\ensuremath{R_{h_2}(\gamma\gamma)}}
\newc{\rtwozz}{\ensuremath{R_{h_2}(ZZ)}}
\newc{\ronegg}{\ensuremath{R_{h_1}(\gamma\gamma)}}
\newc{\ronezz}{\ensuremath{R_{h_1}(ZZ)}}
\newc{\rsiggg}{\ensuremath{R_{h_\textrm{sig}}(\gamma\gamma)}}
\newc{\rsigzz}{\ensuremath{R_{h_\textrm{sig}}(ZZ)}}
\newc{\llbar}{\ensuremath{\ell\bar{\ell}}}
\newc{\tauptaum}{\ensuremath{ \tau^+\tau^-}}
\newc{\qqbar}{\ensuremath{ q\bar{q}}} \newc{\ppbar}{\ensuremath{ p\bar{p}}}
\newc{\bbbar}{\ensuremath{ b\bar{b}}} \newc{\ttbar}{\ensuremath{ t\bar{t}}}
\newc{\ffbar}{\ensuremath{ f\bar{f}}} \newc{\tautaubar}{\ensuremath{ \tau\bar{\tau}}}

\newc{\mchi}{\ensuremath{m_\neutone}}
\newc{\squark}{\ensuremath{\tilde{q}}}
\newc{\slepton}{\ensuremath{\tilde{l}}}
\newc{\gluino}{\ensuremath{\tilde{g}}} 
\newc{\mgluino}{\ensuremath{{m_{\gluino}}}}
\newc{\wino}{\ensuremath{\tilde{W}}} 
\newc{\mwino}{\ensuremath{{m_{\wino}}}}
\newc{\tone}{\ensuremath{{\tilde{t}_1}}}
\newc{\bone}{\ensuremath{{\tilde{b}_1}}}
\newc{\Hone}{\ensuremath{{\tilde{H}_{1}}}}
\newc{\Htwo}{\ensuremath{{\tilde{H}_{2}}}}
\newc{\Hhtwo}{\ensuremath{{H_{2}}}}
\newc{\qli}{\ensuremath{{\tilde{Q}_{i}}}}
\newc{\uri}{\ensuremath{{\tilde{u}_{i}}}}
\newc{\dri}{\ensuremath{{\tilde{d}_{i}}}}
\newc{\lli}{\ensuremath{{\tilde{L}_{i}}}}
\newc{\eri}{\ensuremath{{\tilde{e}_{i}}}}

\newc{\sthw}{\ensuremath{ \sin\theta_W}}              \newc{\cthw}{\ensuremath{\cos\theta_W}}
\newc{\tanthw}{\ensuremath{ \tan\theta_W}}              \newc{\cotthw}{\ensuremath{\cot\theta_W}}
\newc{\ssqthw}{\ensuremath{\sin^2 \theta_W}}
\newc{\msbar}{\ensuremath{\overline{MS}}} \newc{\drbar}{\ensuremath{\overline{DR}}}
\newc{\mtmtsmmsbar}{\ensuremath{ m_t(m_t)^{\msbar}_{{\mathrm{SM}}}}}
\newc{\mtmtsmdrbar}{\ensuremath{ m_t(m_t)^{\drbar}_{{\mathrm{SM}}}}}
\newc{\mtmtmssmdrbar}{\ensuremath{ m_t(m_t)^{\drbar}_{{\mathrm{SUSY}}}}}
\newc{\mbmbmsbar}{\ensuremath{ m_b(m_b)^{\msbar} }}
\newc{\mbmbsmmsbar}{\ensuremath{ m_b(m_b)^{\msbar}_{{\mathrm{SM}}}}}
\newc{\mbmzsmmsbar}{\ensuremath{ m_b(\mz)^{\msbar}_{{\mathrm{SM}}}}}
\newc{\mbmzsmdrbar}{\ensuremath{ m_b(\mz)^{\drbar}_{{\mathrm{SM}}}}}
\newc{\mbmzmssmdrbar}{\ensuremath{ m_b(\mz)^{\drbar}_{{\mathrm{SUSY}}}}}
\newc{\mtaumzsmmsbar}{\ensuremath{ m_{\tau}(\mz)^{\msbar}_{{\mathrm{SM}}}}}
\newc{\mtaumzsmdrbar}{\ensuremath{ m_{\tau}(\mz)^{\drbar}_{{\mathrm{SM}}}}}
\newc{\mtaumzmssmdrbar}{\ensuremath{ m_{\tau}(\mz)^{\drbar}_{{\mathrm{SUSY}}}}}
\newc{\alphasmzms}{\ensuremath{\alpha_s(M_Z)^{\overline{MS}}}}
\newc{\alphaimzms}[1]{\ensuremath{\alpha_{#1}(M_Z)^{\overline{MS}}}}
\newc{\alphaemmz}{\ensuremath{\alpha_{\mathrm{em}}(M_Z)^{\overline{MS}}}}

\newc{\mzero}{\ensuremath{{m_0}}}
\newc{\mhalf}{\ensuremath{ m_{1/2}}}
\newc{\tanb}{\ensuremath{\tan\beta}}
\newc{\azero}{\ensuremath{ A_0}}
\newc{\signmu}{\ensuremath{\rm{sgn}\,\mu}}
\newc{\atau}{\ensuremath{{A_{\tau}}}}
\newc{\mueff}{\ensuremath{\mu_{\rm{eff}}}}
\newc{\lam}{\ensuremath{{\lambda}}}
\newc{\kap}{\ensuremath{{\kappa}}}
\newc{\alam}{\ensuremath{{A_{\lambda}}}}
\newc{\akap}{\ensuremath{{A_{\kappa}}}}
\newc{\hs}{\ensuremath{ H_s}}      
\newc{\mhs}{\ensuremath{ m_{H_s}}} 
\newc{\mgut}{\ensuremath{ M_{\rm GUT}}}
\newc{\gut}{\ensuremath{{\rm GUT}}}
\newc{\mplanck}{\ensuremath{ M_{\rm P}}}      \newc{\mpl}{\ensuremath{ M_{\rm Pl}}}
\newc{\msusy}{\ensuremath{ M_{\rm SUSY}}}      \newc{\ms}{\ensuremath{ M_{\rm S}}}
 \newc{\hu}{\ensuremath{ H_u}}       \newc{\hd}{\ensuremath{ H_d}}
 \newc{\mhu}{\ensuremath{ m_{H_u}}}       \newc{\mhd}{\ensuremath{ m_{H_d}}}
 \newc{\mhuew}{\ensuremath{ m^{\ast}_{H_u}}}       \newc{\mhdew}{\ensuremath{ m^{\ast}_{H_d}}}
 \newc{\mhuewsq}{\ensuremath{ m^{\ast\, 2}_{H_u}}}       \newc{\mhdewsq}{\ensuremath{ m^{\ast\, 2}_{H_d}}}
 \newc{\mhl}{\ensuremath{m_\hl}} 
 \newc{\mhone}{\ensuremath{m_{h_1}}} 
 \newc{\mhtwo}{\ensuremath{m_{h_2}}} 
 \newc{\mhi}{\ensuremath{m_{\tilde{h}}}} 
 \newc{\mul}{\ensuremath{m_{\tilde{u}_L}}} 
 \newc{\mbone}{\ensuremath{m_{\tilde{b}_1}}}  
 \newc{\mtone}{\ensuremath{m_{\tilde{t}_1}}} 
 \newc{\ma}{\ensuremath{m_A}} 
 \newc{\mH}{\ensuremath{m_H}} 
 \newc{\maone}{\ensuremath{m_{a_1}}} 
 \newc{\matwo}{\ensuremath{m_{a_2}}}
 \newc{\hone}{\ensuremath{h_1}}
 \newc{\htwo}{\ensuremath{h_2}}
 \newc{\aone}{\ensuremath{a_1}}
 \newc{\atwo}{\ensuremath{a_2}}
 \newc{\mqthree}{\ensuremath{m_{\tilde{Q}_3}^2}}
 \newc{\muthree}{\ensuremath{m_{\tilde{u}_3}^2}}
 \newc{\mqli}{\ensuremath{m_{\tilde{Q}_{i}}}}
 \newc{\muri}{\ensuremath{m_{\tilde{u}_{i}}}}
 \newc{\mdri}{\ensuremath{m_{\tilde{d}_{i}}}}
 \newc{\mlli}{\ensuremath{m_{\tilde{L}_{i}}}}
 \newc{\meri}{\ensuremath{m_{\tilde{e}_{i}}}}
 \newc{\ts}{\ensuremath{T_{SUSY}}}

\newc{\sigsip}{\ensuremath{\sigma^{\rm SI}_{p}}}	\newc{\sigsin}{\ensuremath{\sigma^{\rm SI}_{n}}}
\newc{\sigsdp}{\ensuremath{\sigma^{\rm SD}_{p}}}	\newc{\sigsdn}{\ensuremath{\sigma^{\rm SD}_{n}}}
\newc{\sigsi}{\ensuremath{\sigma^{\rm SI}}}	\newc{\sigsd}{\ensuremath{\sigma^{\rm SD}}}
\newc{\abund}{\ensuremath{ \Omega h^2}}
\newc{\omegadm}{\ensuremath{ \Omega_{{\rm DM}}}}     \newc{\abunddm}{\ensuremath{ \Omega_{{\rm DM}} h^2}} 
\newc{\omegam}{\ensuremath{ \Omega_{{\rm m}}}}       \newc{\abundm}{\ensuremath{ \Omega_{{\rm m}} h^2}}
\newc{\omegab}{\ensuremath{ \Omega_{{\rm b}}}}	\newc{\abundb}{\ensuremath{ \Omega_{{\rm b}} h^2}}
\newc{\omegatot}{\ensuremath{ \Omega_{{\rm TOT}}}}
\newc{\omegacdm}{\ensuremath{ \Omega_{{\rm CDM}}}}   \newc{\abundcdm}{\ensuremath{ \Omega_{{\rm CDM}} h^2}}
\newc{\omegalambda}{\ensuremath{ \Omega_{\Lambda}}} \newc{\abundlambda}{\ensuremath{ \Omega_{\Lambda} h^2}}
\newc{\omegarad}{\ensuremath{ \Omega_{{\rm rad}}}}  \newc{\abundrad}{\ensuremath{ \Omega_{{\rm rad}} h^2}}
\newc{\rhocrit}{\ensuremath{ \rho_{\rm crit}}}
\newc{\rhochi}{\ensuremath{ \rho_{\chi}}}
\newc{\abunchi}{\ensuremath{\Omega_\chi h^2}}
\newc{\abundlsp}{\ensuremath{\Omega_{\rm LSP}h^2}}

\newc{\amu}{\ensuremath{ a_{\mu}}}        \newc{\amususy}{\ensuremath{ a_{\mu}^{\mathrm{SUSY}}}}
\newc{\amuexpt}{\ensuremath{ a_{\mu}^{\mathrm{expt}}}}        \newc{\amusm}{\ensuremath{ a_{\mu}^{\mathrm{SM}}}}
\newc\deltaamu{\ensuremath{\Delta a_{\mu}}} \newc{\deltaamususy}{\ensuremath{\delta a_{\mu}^{\mathrm{SUSY}}}}
\newc\gmtwo{\ensuremath{ (g-2)_{\mu}}} 
\newc{\deltagmtwomususy}{\ensuremath{\delta\left(g-2\right)_{\mu}^{\mathrm{SUSY}}}}
\newc{\deltagmtwomu}{\ensuremath{\delta\left(g-2\right)_{\mu}}}
\newc\BR{\ensuremath{\rm BR}}
\newc\bsgamma{\ensuremath{ b\rightarrow s \gamma }}
\newc\bxsgamma{\ensuremath{\overline{B}\rightarrow X_{s}\gamma}}
\newc\brbsgamma{\ensuremath{\BR\left(\bsgamma\right)}}
\newc\brbxsgamma{\ensuremath{\BR\left(\bxsgamma\right)}}
\newc\bsmumu{\ensuremath{B_s\to\mu^+\mu^-}}
\newc\brbsmumu{\ensuremath{\BR\left(B_s\to\mu^+\mu^-\right)}}
\newc\bdmmumu{\ensuremath{\overline{B}_d\to\mu^+\mu^-}}
\newc\bbbarmix{\ensuremath{\overline{B}_s\mbox{-}B_s}}      
\newc\delmbs{\ensuremath{\Delta M_{B_s}}}
\newc{\butaunu}{\ensuremath{B_u \rightarrow \tau \nu}}
\newc{\brbutaunu}{\ensuremath{\BR\left(B_u \rightarrow \tau \nu\right)}}

\newcommand*{\reftable}[1]{Table~\ref{#1}}         
       
\newcommand*{\reffig}[1]{Fig.~\ref{#1}}
 
        \newcommand*{\refeq}[1]{Eq.~(\ref{#1})}
     \newcommand*{\refsec}[1]{Sec.~\ref{#1}}

\newcommand*{\neutone}{\ensuremath{\chi^0_1}}

\newcommand*{\spheno}{{\tt SPheno}}

\let\oldcite\cite
\renewcommand*{\cite}{~\oldcite}

\newcommand*{\hl}{\ensuremath{h}}

\newcommand{\Sla}[1]%
{\kern0.12em{\raise.15ex\hbox{$/$}\kern-.74em #1}}

\renewcommand{\Re}{\mathop{\textrm{Re}} }
\renewcommand{\Im}{\mathop{\textrm{Im}} }

\newcommand{\Group}[2]{{ \hbox{{\itshape{#1}}($#2$)} }}
\newcommand{\U}[1]{\Group{U\kern0.05em}{#1}}
\newcommand{\SU}[1]{\Group{SU\kern0.1em}{#1}}
\newcommand{\SL}[1]{\Group{SL\kern0.05em}{#1}}
\newcommand{\Sp}[1]{\Group{Sp\kern0.05em}{#1}}
\newcommand{\SO}[1]{\Group{SO\kern0.1em}{#1}}

\newcommand{\mybar}[1]%
    {{\kern 0.8pt\overline{\kern -0.8pt#1\kern -0.8pt}\kern 0.8pt}}
\newcommand{\sla}[1]%
    {{\raise.15ex\hbox{$/$}\kern-.57em #1}}
\newcommand{\roughly}[1]%
    {{ \mathrel{\raise.3ex\hbox{ $#1$\kern-.75em\lower1ex\hbox{$\sim$}} } }}

\newcommand{\nop}[1]{:\kern-.3em#1\kern-.3em:}

\providecommand{\abs}[1]{\lvert#1\rvert}

\newcommand{\al}{\ensuremath{\alpha}}
\newcommand{\ga}{\ensuremath{\gamma}}

\newcommand{\de}{\ensuremath{\delta}}
\newcommand{\De}{\ensuremath{\Delta}}

\newcommand{\ka}{\ensuremath{\kappa}}
\newcommand{\la}{\ensuremath{\lambda}}

\newcommand{\rh}{\ensuremath{\rho}}
\newcommand{\si}{\ensuremath{\sigma}}

\newcommand{\ta}{\ensuremath{\tau}}

\newcommand*{\mad}{{\tt MadGraph5$\_$aMC@NLO}}

\newcommand{\GeV}{ \ensuremath{\mathrm{~GeV}} }

\newcommand{\hc}{\mathrm{H.c.}} 
\newcommand{\n}{\notag \\}
\newcommand{\mcl}[1]{\mathcal{#1}}

\numberwithin{equation}{section}

\begin{document}

\title{Constraints on charmphilic solutions to the muon $g-2$ with leptoquarks}

\author{Kamila Kowalska, Enrico Maria Sessolo, and Yasuhiro Yamamoto\\[2ex]
\small {\em National Centre for Nuclear Research}\\
\small {\em Ho{\. z}a 69, 00-681 Warsaw, Poland }\\
}
\date{}

{\let\newpage\relax\maketitle}
\centering
\url{kamila.kowalska@ncbj.gov.pl},\\
\url{enrico.sessolo@ncbj.gov.pl},\\
\url{yasuhiro.yamamoto@ncbj.gov.pl}

\abstract{We derive constraints from flavor and LHC searches on charmphilic contributions to the muon anomalous magnetic moment with a single leptoquark. 
Only the scalar leptoquarks $S_1$ and $R_2$ are relevant for the present analysis. 
We find that for $S_1$ some parameter space remains consistent at $2\,\sigma$ with the Brookhaven National Laboratory 
measurement of \gmtwo, 
under the assumption that the left-type coupling between the muon   
and the charm quark is a free parameter. The surviving parameter space
is, on the other hand, going to be probed in its entirety at the LHC with 300\invfb of luminosity or less. 
All other possibilities are excluded by the LHC dimuon search results in combination with several flavor bounds, which
together require one to introduce sizable couplings to the top quark to be evaded.}


\setcounter{footnote}{0}
\section{Introduction\label{sec:intro}}

The currently running Muon g-2 experiment at Fermilab\cite{Grange:2015fou} 
will measure the anomalous magnetic moment of the muon, \gmtwo,  
with precision $0.1\,\textrm{ppm}$. The experiment will be then backed up in the near future 
by another one at J-PARC\cite{Mibe:2010zz}, 
designed to reach a comparable sensitivity with a different experimental setup and thus substantially reduce the impact
of systematic uncertainties. 

These new measurements will play a vital role in constraining the parameter space of the models of new physics invoked in recent years to explain 
the discrepancy of the previous determination of \gmtwo\cite{Bennett:2006fi}, at Brookhaven's BNL, with the Standard Model (SM) expectation. 
When taking into account recent estimates of the hadronic vacuum polarization uncertainties (see, e.g.,\cite{Davier:2016iru,Jegerlehner:2017lbd}), 
the BNL value was found to be in excess of the SM by approximately $3.5\,\sigma$: $\deltagmtwomu=(27.4\pm 7.6)\times 10^{-10}$, according to the estimate\cite{Davier:2016iru}, 
or $\deltagmtwomu=(31.3\pm 7.7)\times 10^{-10}$, according to Ref.\cite{Jegerlehner:2017lbd}.

The Large Hadron Collider (LHC) has provided strong constraints on  
several realistic models that explain the BNL anomaly by introducing
new particles not far above the electroweak symmetry-breaking (EWSB) scale. 
For example, in models based on low-scale supersymmetry\cite{Moroi:1995yh,Cho:2000sf,Martin:2001st} searches for new physics 
with 2--3 leptons in the final state and large missing energy\cite{Aaboud:2018jiw,Aaboud:2017leg,Sirunyan:2017lae,Sirunyan:2018iwl} 
have proven very effective in dramatically restricting the parameter space compatible with \gmtwo\
(see, e.g., Refs.\cite{Endo:2013bba,Akula:2013ioa,Fowlie:2013oua,Endo:2013lva,Chakraborti:2014gea,Kowalska:2015zja,Padley:2015uma} for early LHC studies). 
And, in general, the same constraints apply to virtually any construction where 
a solution to the \gmtwo\ anomaly involves colorless particles and a dark sector\cite{Kowalska:2017iqv}.
Even scenarios where \deltagmtwomu\ is obtained
without directly involving stable invisible particles, so that one cannot rely on missing energy as a handle to discriminate from the background, 
are by now strongly constrained\cite{Dermisek:2013gta,Freitas:2014pua} by the null results of LHC searches with multiple light leptons 
in the final state\cite{ATLAS-CONF-2013-070,Aad:2014hja}.
In general the emerging picture is that one is forced to consider the addition of at least two more states about or above 
the EWSB scale to explain \deltagmtwomu\ and, at the same time, avoid the tightening LHC bounds.

This is not the case for leptoquarks. 
As they can boost the value of the anomalous magnetic moment of the muon by coupling 
with a chirality-flip interaction to a heavy quark, 
they can yield $\deltagmtwomu\sim 10^{-9}$ with acceptable couplings even when their mass is in the 
TeV range\cite{Djouadi:1989md,Chakraverty:2001yg,Cheung:2001ip,Queiroz:2014zfa}. 

Leptoquarks have attracted significant interest
in recent years for several reasons. 
From the theoretical point of view their existence arises for example 
as a natural byproduct of the Grand Unification of fundamental interactions. They are also featured in models of supersymmetry without R-parity
and in some composite models. 
But perhaps the main reason that has rendered them
a current staple in the high-energy physics literature is the fact that 
they provide a plausible solution for the recent LHCb Collaboration flavor anomalies (see, e.g., Refs.\cite{Sakaki:2013bfa,Hiller:2014yaa,Gripaios:2014tna,Varzielas:2015iva,Calibbi:2015kma,Freytsis:2015qca,Bauer:2015knc,Fajfer:2015ycq,Barbieri:2015yvd} 
for early explanations based on leptoquarks)
if they present non-negligible couplings to the third-generation quarks.

It is possibly for this reason that the
most recent studies tackling 
the \gmtwo\ anomaly with leptoquark solutions 
focus mostly on the couplings to the quarks of the third generation\cite{Bauer:2015knc,ColuccioLeskow:2016dox,Bar-Shalom:2018ure}. 
This choice might also present itself as a matter of convenience, as the third generation is 
trivially expected to be less constrained phenomenologically than the second.
Here, however, we move in the opposite direction and focus instead on the case of \textit{charmphilic} leptoquarks, i.e., leptoquarks that produce a signal in \gmtwo\ 
by coupling to the second generation quarks\cite{Cheung:2001ip,Bauer:2015knc}. 
Our motivation lies on the fact that, one should not forget, the flavor structure of leptoquark interactions is still
largely unconstrained on theoretical grounds, so that their eventual couplings to the SM particles 
must be inferred (or, in most cases, excluded) by phenomenological analysis. 
Thus, we apply to leptoquarks the spirit of several recent systematic studies\cite{Kowalska:2017iqv,Freitas:2014pua,Calibbi:2018rzv} that, 
in anticipation of the upcoming data from the Muon g-2 experiment, have confronted renormalizable new physics models consistent with a \gmtwo\ signal
with the most recent data from the LHC and complementary input from flavor, precision, and other experiments. 
In light of the fact that the LHCb anomalies might disappear when further data becomes available,
and thus cease to provide a reason for considering predominantly third-generation couplings, we find it perhaps more natural to start with the analysis of the second generation.

We show that almost all realistic and well-motivated 
charmphilic solutions to the \gmtwo\ anomaly are now excluded by 
a combination of recent bounds from the LHC\cite{Aaboud:2017buh,Sirunyan:2018exx} 
and a handful of data from flavor experiments, 
with the exception of narrow slices of the parameter space 
that will be probed in their entirety with 300\invfb\ of luminosity. 
Thus, if a significant \deltagmtwomu\ deviation from the SM
were observed at Fermilab as well, 
this would mean in this framework 
that leptoquark couplings to the third generation are required. We provide estimates of the minimal size of 
third-generation couplings that is necessary to evade all constraints.  

The paper is organized as follows. In \refsec{sec:gm2} we briefly review the expression for the muon anomalous magnetic moment with one leptoquark at one loop.
In \refsec{sec:mod} we introduce the leptoquark models that yield a solution to the BNL \gmtwo\ anomaly with charmphilic couplings, we specify the assumptions involved, and
we estimate the required coupling size at $2\,\sigma$. In \refsec{sec:flavor} we compute flavor 
and electroweak (EW) precision constraints on the parameter space of the models. We numerically simulate the relevant LHC searches and present our main results in \refsec{sec:results};
and we finally conclude in \refsec{sec:summary}. In Appendix~\ref{app:update} we have updated the full analysis in light of the new 
measurement of the anomalous magnetic moment of the muon at Fermilab\cite{PhysRevLett.126.141801}.

\section{The muon \textit{\textbf{g--2}} in minimal leptoquark models \label{sec:gm2}}

We start by considering the generic Yukawa-type interaction of a scalar leptoquark $S$ with the muon, $\mu$, and one quark or antiquark, $q$:
\be
\mathcal{L}\supset g_s\,\bar{\mu}q S+g_p\,\bar{\mu}\gamma_5 q S+\textrm{H.c.}\label{generlagr}
\ee 
It is then well known that $q$ and $S$ contribute at one loop to the calculation of the anomalous magnetic moment of the muon as\cite{Queiroz:2014zfa}
\begin{multline}
\deltagmtwomu=-\frac{N_c m_{\mu}^2}{8\pi^2 m_S^2}\left\{\frac{m_q}{m_{\mu}}\left(|g_s|^2-|g_p|^2\right)\left[Q_S f_1(r)+Q_q f_2 (r)\right]\right.\\
\left.+2\left(|g_s|^2+|g_p|^2\right)\left[Q_S f_3(r)+Q_q f_4 (r)\right]\right\},\label{genergm2}
\end{multline}
where $m_{\mu}$, $m_q$, $m_S$ are masses of the muon, quark, and leptoquark, respectively;  
$N_c=3$ is a color factor; $Q_S, Q_q$ are the leptoquark and quark electric charges (with the convention that $Q_S+Q_q+Q_{\mu}=0$); and loop functions
of $r=m_q^2/m_S^2$ are given by
\bea
f_1\left(r\right)&=&\frac{1}{2\left(1-r\right)^3}\left(1-r^2+2 r \ln r\right)\\
f_2\left(r\right)&=&\frac{1}{2\left(1-r\right)^3}\left(3-4r+r^2+2\ln r\right)\\
f_3\left(r\right)&=&\frac{1}{12\left(1-r\right)^4}\left(-1+6r-3r^2-2r^3+6r^2\ln r\right)\\
f_4\left(r\right)&=&\frac{1}{12\left(1-r\right)^4}\left(2+3r-6r^2+r^3+6r\ln r\right).
\eea  

In chiral leptoquark models
the relative size of the scalar and pseudoscalar couplings, $g_s$, $g_p$, parametrizes the strength of the couplings to the 
left- and right-chiral component of the muon.  
We thus redefine $g_L=g_s+g_p$ and $g_R=(-1)^{\gamma}(g_s-g_p)$, 
where $\gamma=0$ in cases where $q$ in \refeq{generlagr} has the quantum numbers of a SM quark, 
and $\gamma=1$ if gauge invariance requires $q\rightarrow q^c$. 
Equation~(\ref{genergm2}) then becomes, for each generation,
\begin{multline}
\deltagmtwomu=-\frac{N_c m_{\mu}^2}{8\pi^2 M_S^2}\left\{\frac{m_q}{m_{\mu}}\left(-1\right)^{\gamma}\Re\left(g_L g_R^{\ast}\right)\left[Q_S f_1(r)+Q_q f_2 (r)\right]\right.\\
\left.+\left(|g_L|^2+|g_R|^2\right)\left[Q_S f_3(r)+Q_q f_4 (r)\right]\right\},\label{chiralgm2}
\end{multline}
which features explicitly in the first line the chirality-flip term proportional to the quark mass,
generated in those models where leptoquarks 
couple simultaneously to both muon chiral states.   
As was mentioned in \refsec{sec:intro}, by virtue of this coupling
the \gmtwo\ anomaly can be resolved in the presence of leptoquarks at the TeV scale, 
if the leptoquark is coupled to second- or third-generation quarks, $q=c,t$. 
This work focuses on the second generation, as we restrict ourselves to charmphilic Yukawa textures that 
induce a  deviation from the SM value parametrized by $\sim m_c/m_{\mu}\ln(m_S/m_c)$.

Before we proceed to introducing the form of plausible models, 
let us recall that, in a similar fashion to the scalar case, a vector leptoquark $V^{\rho}$ of mass $m_V$ can also contribute to \gmtwo.  
Given a generic coupling to one quark and the muon,
\be
\mathcal{L}\supset g_v V_{\rho}\,\bar{q}\gamma^{\rho}\mu+g_a V_{\rho}\,\bar{q}\gamma^{\rho}\gamma_5\mu+\textrm{H.c.}\,,
\ee
the one-loop contribution to \gmtwo\ in the limit $m_q\ll m_V$ reads, for each generation,
\begin{multline}
\deltagmtwomu=\frac{N_c m_{\mu}^2}{8\pi^2 m_V^2}\left[\left(|g_v|^2+|g_a|^2\right)\left(-\frac{4}{3}Q_q+\frac{5}{3}Q_V\right)\right.\\
\left.+\left(|g_v|^2-|g_a|^2\right)\left(Q_q-Q_V\right)\frac{2 m_q}{m_{\mu}}\right],\label{vectgm2}
\end{multline}
with the convention that $Q_V+Q_q+Q_{\mu}=0$\cite{Queiroz:2014zfa}. 

We will briefly come back to the vector leptoquark case in \refsec{sec:mod}. Let us just point out for now that, unlike scalar leptoquark, 
vector leptoquarks do not induce at one loop a $\sim\ln (m_V/m_q)$ enhancement to the \gmtwo\ value, so that they require very large 
couplings to the second-generation quarks if one wishes to accommodate the BNL measurement.

\section{The models \label{sec:mod}}

\textit{Scalar leptoquarks.} We begin with scalar leptoquarks. There exist 
two sole single-leptoquark cases that lead to a mass-enhanced contribution to \gmtwo\ by coupling to both muon chiral states:
SU(2)$_L$ singlet $S_1$, and SU(2)$_L$ doublet $R_2$\cite{Cheung:2001ip}.\smallskip

\begin{table}[t]
	\begin{center}
		\begin{tabular}{c|c|c|c}
			\hline
			\hline
			\rule{0pt}{2.5ex} \small{} Field  &    SU(3)   &  SU(2)$_L$ & U(1)$_Y$      \\
			\hline
			\hline
			\rule{0pt}{2.5ex} $L_i'$ &  $\mathbf{1}$  & $\mathbf{2}$ & $-1/2$ \\
			\rule{0pt}{2.5ex} $e_{R\,i}'$ &  $\mathbf{1}$  & $\mathbf{1}$ & $1$ \\
			\rule{0pt}{2.5ex} $Q_i'$ &  $\mathbf{3}$  & $\mathbf{2}$ & $1/6$  \\
			\rule{0pt}{2.5ex} $u_{R\,i}'$ &  $\mathbf{\bar{3}}$  & $\mathbf{1}$ & $-2/3$  \\
			\rule{0pt}{2.5ex} $d_{R\,i}'$ &  $\mathbf{\bar{3}}$  & $\mathbf{1}$ & $1/3$ \\
			\rule{0pt}{2.5ex} $H$ &  $\mathbf{1}$  & $\mathbf{2}$ & $1/2$  \\
			\hline
			\rule{0pt}{2.5ex} $S_1$ &  $\mathbf{\bar{3}}$  & $\mathbf{1}$ & $1/3$ \\
			\rule{0pt}{2.5ex} $R_2$ &  $\mathbf{3}$  & $\mathbf{2}$ & $7/6$ \\
			\rule{0pt}{2.5ex} $U_1^{\rho}$ &  $\mathbf{3}$  & $\mathbf{1}$ & $2/3$ \\
			\hline
			\hline
		\end{tabular}
		\caption{Gauge quantum numbers of the SM fields, with generation index $i=1,2,3$, and leptoquarks we consider in this study. }
		\label{tab:qn} 
	\end{center}
\end{table}

\noindent \textbf{Model 1.} Leptoquark $S_1$ is characterized by the SM quantum numbers
\be
S_1:(\mathbf{\bar{3}},\mathbf{1},1/3)\,. 
\ee

Using the Weyl spinor notation, we introduce $CP$-conserving Yukawa-type couplings to the Lagrangian,
\be
\mathcal{L}\supset Y^D_{ij}Q_i'^T(-i\sigma_2)L_j' S_1+Y^S_{ij} u_{R\,i}'^{\ast}e_{R\,j}'^{\ast} S_1+\textrm{H.c.}\,,
\ee
where primed fields are given in the gauge basis, and a sum over the SM generation indices $i,j$ is intended. 
The quantum numbers of the SM fields and of the leptoquarks considered in this work 
are summarized in \reftable{tab:qn}.

Note that the quantum number can allow one to write down Lagrangian 
terms of the form $\mathcal{L}\supset \lam^D_{ij}\,Q_i' Q_j' S_1^{\ast}$ + $\lam^S_{ij}\,u_{R\,i}' d_{R\,j}' S_1$ + H.c., which can lead to fast proton decay. 
In order to forbid these dangerous terms, we assume the existence of a symmetry (for example, 
conservation of baryon and/or lepton number).

After EWSB one can rotate the Lagrangian to the quark mass basis and write down the couplings to the second-generation leptons, required for \gmtwo:
\be
\mathcal{L}\supset \left(-\widetilde{Y}^L_i u_{L\,i}\,\mu_L+\hat{Y}^L_i d_{L\,i}\,\nu_{\mu}+Y^R_i u_{R\,i}^{\ast}\,\mu_R^{\ast}\right)S_1+\textrm{H.c.}\,,\label{mod1secgen}
\ee
where nonprimed fields indicate mass eigenstates, we have defined $Q_i^T\equiv (u_{L\,i},d_{L\,i})$, 
$L_i\equiv (\nu_{i},e_{L\,i})^T$,
and the $L$-type couplings are related to each other via the Cabibbo-Kobayashi-Maskawa (CKM) matrix, $\widetilde{Y}^L_i=\hat{Y}^L_k (V_{\textrm{CKM}}^{\dag})_{ki}$. 

We now implement the charmphilic assumption for the $Y^R_i$ Yukawa couplings by imposing
$Y^R_1=Y^R_3= 0$. The contribution to \gmtwo\ is read off from \refeq{chiralgm2}, where $\gamma=1$, $g_L=-\widetilde{Y}_2^L$, $g_R=Y_2^R$, $m_q=m_c$, 
$Q_q=2/3$, and $Q_S=1/3$. It leads to the $2\,\sigma$ bound  (we use the estimate in Ref.\cite{Davier:2016iru})
\be
3.1\times 10^{-2}\left(\frac{m_{S_1}}{\tev}\right)^2\leq \Re{(\widetilde{Y}^L_2 Y_2^{R\ast})}\left(1+0.17 \ln\frac{m_{S_1}}{\tev}\right)\leq 11\times 10^{-2}\left(\frac{m_{S_1}}{\tev}\right)^2.\label{g2boundMod1}
\ee

For the $L$-type couplings, which are related by gauge invariance through the CKM matrix, 
we base our phenomenological analysis on two limiting cases:
the \textbf{up origin}, featuring $\widetilde{Y}^L_1=\widetilde{Y}^L_3= 0$;
and the \textbf{down origin}, featuring $\hat{Y}^L_1=\hat{Y}^L_3= 0$.

In the up-origin case the couplings $\hat{Y}_1^L$,  $\hat{Y}_2^L$, and $\hat{Y}_3^L$ to the down, strange, and bottom quark, respectively, 
will be generated by multiplication with the CKM matrix. 
As a consequence, the $L$-type coupling $\widetilde{Y}_2^L$ becomes subject
to flavor bounds from, e.g., the $K^+\rightarrow \pi^+\nu\bar{\nu}$ rare decay. On the other hand, in the down-origin case, 
the couplings $\widetilde{Y}_1^L$, $\widetilde{Y}_2^L$, and $\widetilde{Y}_3^L$ to the up, charm, and top quark, respectively, are CKM-generated
and thus the $L$-type coupling $\hat{Y}_2^L$ can be bounded by processes like $D^0\rightarrow \mu^+\mu^-$. 
Flavor-changing neutral current (FCNC) 
constraints do not apply, however, to the charmphilic $R$-type coupling $Y_2^R$. 
If it is required to be large in order to satisfy \refeq{g2boundMod1}, it can be probed directly by collider searches at the LHC.\smallskip

\noindent \textbf{Model 2.} Leptoquark $R_2$ is characterized by the SM quantum numbers
\be
R_2:(\mathbf{3},\mathbf{2},7/6)\,. 
\ee
 
The gauge-invariant Yukawa interactions read in this case 
\be
\mathcal{L}\supset Y^L_{ij}L'^T_i(-i\sigma_2)R_2 u_{R\,j}'+Y^R_{ij}Q'^{\dag}_i R_2 e_{R\,j}'^{\ast}+\textrm{H.c.}\,,
\ee
where a sum over family indices is intended.

In the quark mass basis the couplings to the second-generation leptons are then given by
\be
\mathcal{L}\supset Y^L_{i}u_{R\,i}\left[\mu_L\,s_{(5/3)}-\nu_{\mu}\,s_{(2/3)}\right]+\widetilde{Y}^R_i u_{L\,i}^{\ast}\,\mu_R^{\ast}\,s_{(5/3)}
+\hat{Y}^R_i d_{L\,i}^{\ast}\,\mu_R^{\ast}\,s_{(2/3)}+\textrm{H.c.}\,,\label{mod2secgen}
\ee
where the scalar fields $s_{(5/3)}$ and $s_{(2/3)}$ belong to the $R_2$ doublet (electric charge in parentheses), 
and again we work under the charmphilic assumption,  $Y^L_1=Y^L_3= 0$.

The contribution to \gmtwo\ in Model~2 is given in \refeq{chiralgm2}, where $\gamma=0$, $g_L=Y_2^L$, $g_R=\widetilde{Y}_2^R$, $m_q=m_c$, 
$Q_q=-2/3$, and $Q_S=5/3$. It leads to the $2\,\sigma$ bound
\be
2.7\times 10^{-2}\left(\frac{m_{R_2}}{\tev}\right)^2\leq -\Re{(\widetilde{Y}^R_2 Y_2^{L\ast})}\left(1+0.15 \ln\frac{m_{R_2}}{\tev}\right)\leq 9.5\times 10^{-2}\left(\frac{m_{R_2}}{\tev}\right)^2.\label{g2boundMod2}
\ee

We introduce for the $R$-type couplings an up-origin and a down-origin scenario, 
in analogy to Model~1. In the up-origin case, we generate couplings to the down and bottom quarks, which render the 
$R$-type coupling $\widetilde{Y}_2^R$ subject to constraints, e.g., from the decay $K_L\rightarrow \mu^+\mu^-$. In the down-origin case,
generated $R$-type couplings to the up-type quarks can be constrained by the measurement of the $D^0\rightarrow \mu^+\mu^-$ transition.\smallskip

\textit{Vector leptoquarks.} We finally tackle the case of a vector leptoquark at the TeV scale as the primary responsible for the \gmtwo\ anomaly.
In general, an appropriate treatment of vector states cannot be carried out without some starting assumption on the nature of the UV completion
that gives rise to the leptoquark itself. In fact, the UV completion might produce additional states lighter than the vector leptoquark (as happens, e.g.,
in ``composite'' models\cite{Eichten:1986eq,Lane:1991qh}), 
which, when accounted for, can reduce the relevance of the heavier vector for the computation of the observable in question.
If, on the other hand, the vector leptoquark is a gauge boson, the appropriate treatment of gauge anomalies should be factored in. 

In recent years, several studies\cite{Assad:2017iib,DiLuzio:2017vat,Calibbi:2017qbu,Bordone:2017bld,Fornal:2018dqn} on 
UV completions based on Pati-Salam constructions have pointed out
that an SU(2)$_L$ singlet vector leptoquark $U_1^{\rho}$ can emerge as the lightest state of the new physics spectrum around the TeV scale.
Because of its gauge quantum numbers, however, 
a leptoquark $U_1^{\rho}:(\mathbf{3},\mathbf{1},2/3)$ coupled to second-generation 
quarks does not produce a mass-enhanced contribution to \gmtwo\
with the charm quark in the loop, but rather with the strange quark.

After rotating to the mass basis the Lagrangian, in fact, reads
\be
\mathcal{L}\supset \widetilde{g}^L_i u_{L\,i}^{\ast}U_{1\rho}\,\bar{\sigma}^{\rho}\nu_{\mu}+\hat{g}^L_i d_{L\,i}^{\ast}U_{1\rho}\,\bar{\sigma}^{\rho}\mu_L+
g^R_i\,d_{R\,i} U_{1\rho}\,\sigma^{\rho}\mu_{R}^{\ast}+\textrm{H.c.}\,,\label{mod3secgen}
\ee
in terms of generic gauge couplings $\widetilde{g}_i^L$, $\hat{g}_i^L$, $g_i^R$.
The contribution to \gmtwo\ is obtained from the sum over 3 generations in \refeq{vectgm2} after the following definitions: $g_{v\,1,3}=-g_{a\,1,3}=\hat{g}_{1,3}^L/2$,
$g_{v\,2}=(g_2^R+\hat{g}_2^L)/2$, $g_{a\,2}=(g_2^R-\hat{g}_2^L)/2$, $m_q=m_{d,s,b}$, 
$Q_q=1/3$, and $Q_V=2/3$. 

Since the contribution to \gmtwo\ is not subject to a $\ln(m_V/m_q)$ enhancement and, additionally,
$m_s/m_{\mu}\approx 0.9$, upholding the $2\,\sigma$ bound from \deltagmtwomu\ requires either small mass or significant couplings:
\be
\left|g_i^L\right|=\left|g_i^R\right|\gsim 2.0 \left(\frac{m_{U_1^{\rho}}}{\tev}\right),\label{vecg2bound}
\ee
where we have indicated all $L$-type couplings generically with $g_i^L$.

The current pair-production LHC mass bounds\cite{Sirunyan:2018kzh}, recast for the case of the $U_1^{\rho}$ leptoquark\cite{Schmaltz:2018nls}, yield $m_{U_1^{\rho}}>1.5\tev$. 
Relation~(\ref{vecg2bound}) thus implies that, in order to explain the \gmtwo\ anomaly, strangephilic vector leptoquarks must have coupling size of about 3 or greater. 
Recent LHC dimuon analyses, to which we will come back in \refsec{sec:results}, when applied specifically to this case have 
shown\cite{Angelescu:2018tyl,Schmaltz:2018nls} that a coupling to the strange quark of size 3 or more is excluded. 
Therefore, we will not consider the vector case any further in this work.

\section{Flavor and electroweak precision constraints\label{sec:flavor}}

Due to their \textit{a priori} unconstrained flavor structure, leptoquarks can generate sizable contributions in some flavor observables, which in turn can lead to undesirable flavor signals.
After integrating out the leptoquarks, the models introduced in \refsec{sec:mod}  produce the tree-level effective Lagrangians
\begin{align}
 \mcl{L}_{S1}^\text{eff} =
   \frac{1}{2m_{S1}^2}& \biggl\{
     \widetilde{Y}_{ij}^L   \widetilde{Y}_{kl}^{L\ast} (\bar{u}_{Li}\ga^\mu u_{Lk}) (\bar{e}_{Lj}\ga_\mu e_{Ll}) 
    +\hat{Y}_{ij}^L   \hat{Y}_{kl}^{L\ast} (\bar{d}_{Li}\ga^\mu d_{Lk}) (\bar{\nu}_{Lj}\ga_\mu \nu_{Ll}) \n &
		+Y_{ij}^R   Y_{kl}^{R\ast}(\bar{u}_{Ri}\ga^\mu u_{Rk}) (\bar{e}_{Rj}\ga_\mu e_{Rl})
		+\widetilde{Y}_{ij}^L   \hat{Y}_{kl}^{L\ast}(\bar{u}_{Li}\ga^\mu d_{Lk}) (\bar{e}_{Lj}\ga_\mu \nu_{Ll}) \n&
		+\frac{\widetilde{Y}_{ij}^L   Y_{kl}^{R\ast}}{4} \left[ -4(\bar{u}_{Li} u_{Rk}) (\bar{e}_{Lj} e_{Rl}) +(\bar{u}_{Li}\si^{\mu\nu} u_{Rk}) (\bar{e}_{Lj}\si_{\mu\nu} e_{Rl}) \right] \n &
		+\frac{\hat{Y}_{ij}^L   Y_{kl}^{R\ast}}{4} \left[ -4(\bar{d}_{Li} u_{Rk}) (\bar{\nu}_{Lj} e_{Rl}) +(\bar{d}_{Li}\si^{\mu\nu} u_{Rk}) (\bar{\nu}_{Lj}\si_{\mu\nu}  e_{Rl}) \right]
		+\cdots
	 \biggr\}, \label{efflagrS1}
\end{align}
\begin{align}
 \mcl{L}^\text{eff}_{R2} =
  -\frac{1}{2m_{s_{(5/3)}}^2}& \biggl\{
    \widetilde{Y}_{ij}^R \widetilde{Y}_{kl}^{R\ast} ( \bar{u}_{Li}\ga^\mu u_{Lk} )( \bar{e}_{Rl}\ga_\mu e_{Rj} )
	 + Y_{ij}^L Y_{kl}^{L\ast} ( \bar{u}_{Ri} \ga^\mu u_{Rk} )( \bar{e}_{Ll}\ga_\mu e_{Lj} ) \n &
   +\frac{Y_{ij}^L \widetilde{Y}_{kl}^{R\ast}}{4} \left[
	   4( \bar{u}_{Ri} u_{Lk} )( \bar{e}_{Rl} e_{Lj} ) +(\bar{u}_{Ri}\si_{\mu\nu} u_{Lk} )( \bar{e}_{Rl}\si^{\mu\nu} e_{Lj} )
	 \right] +\cdots
  \biggr\} \n 
  -\frac{1}{2m_{s_{(2/3)}}^2}& \biggl\{
    \hat{Y}_{ij}^R \hat{Y}_{kl}^{R\ast} ( \bar{d}_{Li}\ga^\mu d_{Lk} )( \bar{e}_{Rl}\ga_\mu e_{Rj} )
	 +Y^L_{ij} Y^{L\ast}_{kl} ( \bar{u}_{Ri} \ga^\mu u_{Rk} )( \bar{\nu}_{Ll}\ga_\mu \nu_{Lj} ) \n &
   -\frac{Y_{ij}^L \hat{Y}_{kl}^{R\ast}}{4} \left[
	   4(\bar{u}_{Ri} d_{Lk} )( \bar{e}_{Rl} \nu_{Lj} ) +( \bar{u}_{Ri}\si_{\mu\nu} d_{Lk} )( \bar{e}_{Rl}\si^{\mu\nu} \nu_{Lj} )
	 \right] +\cdots
  \biggr\},\label{efflagrR2}
\end{align}
where the ellipsis indicate the Hermitian conjugate of the shown operators 
if they are not self-conjugate, and, with a slight abuse of notation, we generalize the couplings
of \refsec{sec:mod} to matrices such that $\widetilde{Y}_i^L\equiv \widetilde{Y}_{i2}^L$ and so on.
Since the $\widetilde{Y}_{ij}^{L,R}$ and $\hat{Y}_{ij}^{L,R}$ Yukawa couplings are related to each other via the CKM matrix, 
some FCNCs are inevitably induced at the tree level.

Leptoquark $R_2$ belongs to a doublet representation and consists of two colored scalars, $s_{(5/3)}$ and $s_{(2/3)}$.
Their Yukawa couplings are related to each other, but in principle experimental constraints on the couplings depend on the scalars' individual masses.			
The mass difference of the doublet states violates the custodial symmetry, so that the $T$-parameter is sensitive to it.
The leptoquark contribution to the $T$-parameter reads
\begin{align}
 \De T =&
   \frac{3}{16\pi m_Z^2 s_W^2 c_W^2} \left( m_{(5/3)}^2+m_{(2/3)}^2 -\frac{2m_{(5/3)}^2 m_{(2/3)}^2}{m_{(5/3)}^2-m_{(2/3)}^2}\ln\frac{m_{(5/3)}^2}{m_{(2/3)}^2} \right)\\
 =&
   \frac{\De m^2}{4\pi m_Z^2 s_W^2 c_W^2} \left[ 1+O\left( \frac{\De m^2}{m_{(2/3)}^2} \right) \right],
\end{align}
where $m_Z$ is the $Z$ boson mass, $s_W,\,c_W$ are the sine and cosine of the Weinberg angle, and 
$\De m =m_{(5/3)} -m_{(2/3)}$, see Ref.\cite{Keith:1997fv}.

Given the observed limit, $\De T = 0.09\pm 0.13$\cite{Baak:2014ora}, at $1\,\sigma$ the mass difference must be bounded by
\begin{align}
  \abs{\De m} =2s_W c_W \sqrt{\pi \abs{\De T}} m_Z \lesssim 63\GeV.
\end{align}
Since the typical mass region studied in this paper is a few TeV, we can easily neglect this difference, given the bound on $\Delta T$. 
Thus, we perform the analysis in this paper in the approximation where the two states are degenerate.

\subsection{Down origin} 

The down-origin ansatz makes a solution for the \gmtwo\ anomaly via charmphilic leptoquark 
inconsistent with the measurement of the branching ratio of the rare flavor process $D^0\rightarrow \mu^+\mu^-$\cite{deBoer:2015boa,Bauer:2015knc}.

The most recent measurement\cite{Aaij:2013cza} of the branching ratio at LHCb reads, at the 95\%~C.L., 
\be
\textrm{Br}(D^0\rightarrow \mu^+\mu^-)<7.6\times 10^{-9}.\label{doexp}
\ee 
We write the branching ratio in the most general form as\cite{Fajfer:2015mia,Cai:2017wry}
\begin{multline}
  \textrm{Br}\left(D^0\to\mu^+\mu^-\right) = \tau_D \frac{f_D^2 m_D^3 }{256\pi} \frac{m_D^2}{m_c^2}
  \left[
    \left|C^D_{SRR}-C^D_{SLL}\right|^2\right.\\
\left. +\left| C^D_{SRR}+C^D_{SLL} -\frac{2m_\mu m_c}{m_D^2}\left(C^D_{VLL}+C^D_{VRR}-C^D_{VRL}-C^D_{VLR}\right)\right|^2
	\right],\label{brD0}
\end{multline}
where the parameters relative to $D$ meson properties are featured in \reftable{TabD}. The SM contribution is dominated by the long-distance effect which is smaller than 
$10^{-12}$\cite{Paul:2010pq}.

\begin{table}
\centering
\begin{tabular}{|c|c|}
\hline 
 $\hbar$ & 6.58 $\times 10^{-25}$\gev sec. \\
 $m_\mu$ & 106\mev \\
   $\la$ & 0.226 \\
\hline
\end{tabular}
\qquad
\begin{tabular}{|c|c|}
\hline 
 $\ta_D$ & 4.10 $\times 10^{-13}$ sec. \\
   $m_D$ & 1.86\gev\\
   $f_D$ &  212\mev \\
   $m_c$ & 1.28\gev \\
 \hline
\end{tabular}
\caption{
Numerical parameters relative to the calculation of $\textrm{Br}(D^0\to\mu^+\mu^-)$, \refeq{brD0}.
 We take the muon mass and $\hbar$ as in Ref.\cite{Mohr:2015ccw}, whereas
 other constants are set at their Particle Data Group (PDG) value\cite{Tanabashi:2018oca}.
}
\label{TabD}
\end{table}

As a function of the couplings of Model~1, the Wilson coefficients can be expressed as
\begin{align}
  C_{VLL}^D =& -\frac{1}{2m_{S1}^2} \widetilde{Y}^L_1 \widetilde{Y}^{L\ast}_2
	          =  -\frac{1}{2m_{S1}^2} \frac{\la}{1-\la^2/2} \left|\widetilde{Y}^L_2 \right|^2 , \\
  C_{VRR}^D =& -\frac{1}{2m_{S1}^2} Y^R_1 Y^{R\ast}_2, \\
  C_{SLL}^D =&  \frac{\eta_{\textrm{QCD}}}{2m_{S1}^2} \widetilde{Y}^L_1 Y_2^{R\ast}
	          =   \frac{\eta_{\textrm{QCD}}}{2m_{S1}^2} \frac{\la}{1-\la^2/2} \widetilde{Y}^L_2 Y_2^{R\ast}, \\
  C_{SRR}^D =&  \frac{\eta_{\textrm{QCD}}}{2m_{S1}^2} Y_1^R \widetilde{Y}_2^{L\ast}, 
\end{align}
where $\eta_{\textrm{QCD}}$ stands for the QCD running effect of the scalar operators, $C_{SXX}(m_c)/$ $C_{SXX}(m_{S_1})$, and 
we used the fact that, in terms of the Wolfenstein parameters, 
$\widetilde{Y}_1^L = \lam \hat{Y}_2^L$ and $\widetilde{Y}_2^L =(1- \lam^2/2) \hat{Y}^L_2$.
Since $\eta_{\textrm{QCD}}=1.99$ for $m_{S_1}=1.5\TeV$\cite{Cai:2017wry},
we assume $\eta_{\textrm{QCD}}\approx 2$ in the following analysis.

Recall from \refsec{sec:mod} that, under the charmphilic ansatz, Yukawa coupling $Y_1^R$ is identically set to zero, so that \refeq{doexp} can be used to derive 
an upper bound on the product $\widetilde{Y}_2^L Y_2^{R\ast}$.
At this point, however, one could wonder whether the same bound could be relaxed for specific nonzero values of $Y_1^R$, 
due to a cancellation between different Wilson coefficients. To derive the bound, it is thus worth treating the coupling $Y_1^R$, 
which does not contribute to \gmtwo, as a free parameter.

To analyze  the impact of $Y_1^R$, we calculate the minimum of the branching ratio with respect to $Y_1^R$ and obtain
\begin{align}
 \text{Br}\left(D^0\to\mu^+\mu^-\right)_{\textrm{min}} = \frac{5.4\times 10^{-2}}{\pi} \ta_D m_D \left( \frac{f_D m_D^2 }{32m_c m_{S1}^2} \left| \widetilde{Y}_2^L Y_2^{R\ast}\right| \right)^2,
\end{align}
under the assumption that $|\widetilde{Y}^L_2/Y^R_2|\ll 1$. The above equation thus yields a conservative
bound on the product of the charmphilic couplings, which reads
\begin{align}
\left|\widetilde{Y}^L_2 Y^{R\ast}_2\right| \leq 1.1\times 10^{-2}\left(\frac{m_{S_1}}{\tev}\right)^2.\label{D0bound}
\end{align}

Note that, 
since the rare $D$ decay bound constrains the absolute value of the Yukawa coupling product, it 
cannot be avoided by introducing the imaginary parts of the couplings.
By comparing \refeq{D0bound} and \refeq{g2boundMod1}, we see that $D^0$ decay excludes the entire $2\,\sigma$ 
region of \deltagmtwomu\ for the down-origin scenario of Model~1. 
A calculation along similar lines, with $\widetilde{Y}^L_i\leftrightarrow \widetilde{Y}^R_i$,  
$\hat{Y}^L_i\leftrightarrow \hat{Y}^R_i$, $Y^R_i\leftrightarrow Y^L_i$ in \refeq{brD0}, leads to the exclusion of the down-origin case in  
Model~2.

One could further wonder, at this point, about how large an eventual coupling to the third generation should be   
in order to avoid full exclusion of the charmphilic scenario.
We can evaluate this for Model~1 
by recalling that, in the presence of the third generation, \refeq{g2boundMod1} is modified into
 \be
3.1\times 10^{-2}\left(\frac{m_{S_1}}{\tev}\right)^2\leq a_c \Re{(\widetilde{Y}^L_2 Y_2^{R\ast})}+20.7\,a_t \Re{(\widetilde{Y}^L_3 Y_3^{R\ast})}\leq 11\times 10^{-2}\left(\frac{m_{S_1}}{\tev}\right)^2,\label{g2modified}
\ee
where $a_c=1+0.17 \ln(m_{S_1}/\tev)$ and $a_t=1+1.06 \ln(m_{S_1}/\tev)$\cite{Bauer:2015knc}.
Equation~(\ref{D0bound}) then implies
\be
\Re{(\widetilde{Y}^L_3 Y_3^{R\ast})}\gsim \frac{1-0.09\ln(m_{S_1}/\tev)}{1+1.06\ln(m_{S_1}/\tev)}\times 1.0 \times 10^{-3} 
\left(\frac{m_{S_1}}{\tev}\right)^2.
\ee

\begin{figure}[t]
\centering
\includegraphics[width=0.60\textwidth]{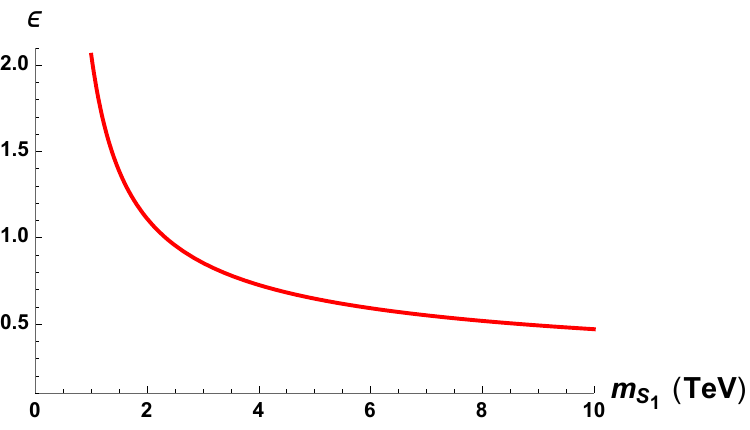}
\caption{The minimum value of $\epsilon=Y_3^R/Y_2^R$, as a function of mass $m_{S_1}$, required in Model~1 (down-origin case) 
to obtain the correct \deltagmtwomu\ and at the same time respect the 95\%~C.L. $\textrm{Br}(D^0\to\mu^+\mu^-)$ bound.}
\label{fig:top_phil_eps}
\end{figure}

If seeking to obtain $\Re(\widetilde{Y}^L_3 Y_3^{R\ast})$ in the simplest possible way, by generating a small contribution 
$Y_3^R=-\epsilon Y_2^R$, one must require $\epsilon\sim\mathcal{O}(1)$ to invalidate the strong 
$D^0$ constraint on the second-generation couplings. We plot in \reffig{fig:top_phil_eps} the leptoquark-mass dependence of this minimally required $\epsilon$.

An equivalent calculation shows that in Model~2 one gets
\be
\Re(\widetilde{Y}^R_3 Y_3^{L\ast})\gsim \frac{1-0.10\ln(m_{R_2}/\tev)}{1+0.61\ln(m_{R_2}/\tev)}\times 0.48 \times 10^{-3}\label{topphil} 
\left(\frac{m_{R_2}}{\tev}\right)^2,
\ee
which does not change the size of the required $\epsilon$ by much.

We point out finally that the measurement of $\textrm{Br}(D^0\to \mu^+\mu^-)$ at LHCb, \refeq{doexp}, is already a few years old, and as such it is probably on track to be renewed with 
fresh data in the near future. If a new determination tightens the upper bound with respect to \refeq{doexp}, then the minimally required top coupling 
is going to be even larger.\footnote{The fact that we need third-generation couplings to explain \deltagmtwomu\ once the bound~(\ref{D0bound}) is enforced
raises the question of whether the scenarios investigated here can be made consistent with the recently observed flavor anomalies at BABAR, Belle, and LHCb, 
in particular the branching ratio measurements providing tantalizing hints of lepton-flavor nonuniversality\cite{Lees:2013uzd,Aaij:2015yra,Hirose:2016wfn,Aaij:2014ora,Aaij:2017vbb}. 
Explanations of the observed excesses in $R_{D^{(\ast)}}=\textrm{Br}(\overline{B}\to D^{(\ast)}\tau\nu_{\tau})/\textrm{Br}(\overline{B}\to D^{(\ast)}l \nu_{l})$ based on $S_1$ and $R_2$ 
seem to imply nonzero couplings between the leptoquark and the $\tau$ ($\nu_{\tau}$)~(see, e.g., Refs.\cite{Sakaki:2013bfa,Bauer:2015knc,Cai:2017wry,Angelescu:2018tyl}), 
and thus lie outside of the main focus of this work. On the other hand, the deficits with respect to the SM 
observed in the $R_{K}$ and $R_{K^{\ast}}$ ratios can admit a viable explanation with couplings to the muon only. For example, the best-fit point to global analyses has been shown to imply, in Model~1, 
$m_{S_1}=5.2\tev$, $\hat{Y}_2^L=0.15$,  $\hat{Y}_3^L\approx -\sqrt{4\pi}$, with $Y_3^R \approx Y_2^R$ very small or close to zero\cite{Angelescu:2018tyl}. 
We find that this is consistent with \deltagmtwomu\ at $2\,\sigma$, provided  $|Y_3^R| \approx | Y_2^R| \gsim 0.005$. (Note, incidentally,
that this can be regarded as a  ``topphilic'' solution and thus a deeper investigation of its properties also somewhat exceeds the purpose of this paper.)  
The current LHC bounds from rare top decays are still unconstraining,
implying $|\hat{Y}_2^L \hat{Y}_3^L |\lesssim \mathcal{O}(1)\times(m_{S_1}/\tev)^2$\cite{Durieux:2014xla,Chala:2018agk}.    
Conversely, the best-fit solution in Model~2 
requires enforcing $\hat{Y}_{i=1,2,3}^R\approx 0$ to very high precision, 
in order to suppress the unwanted tree-level contributions to the Wilson coefficients $C_9$ and $C_{10}$, 
which would lead to $R_{K^{(\ast)}}>R_{K^{(\ast)}}^{\textrm{SM}}$\cite{Becirevic:2017jtw}. 
In the limit of zero right-handed couplings, the anomaly in \gmtwo\ cannot be explained.}

\subsection{Up origin} 

Flavor constraints provide a less clear-cut picture in the case of the up-origin Yukawa texture.
As can be inferred from the form of Eqs.~(\ref{mod1secgen}) and~(\ref{mod2secgen}), in this case
different flavor processes place bounds on the couplings of Model~1 and Model~2 separately.

In Model~1, the strongest bound can be derived from the measurement of the branching ratio $\textrm{Br}(K^+\to\pi^+\nu\bar{\nu})$\cite{Buras:2014fpa,Kumar:2016omp}.
The observed value is $\text{Br}(K^+\to\pi^+\nu\bar{\nu})=(1.73^{+1.15}_{-1.05})\times 10^{-10}$, 
which was measured by the E949 Collaboration\cite{Artamonov:2008qb}.

The effective Lagrangian relevant to this process can be written down as\cite{Kumar:2016omp}
\begin{align}
 \mcl{L}_{K\pi\nu\nu} = \sum_{\ell=e,\mu,\ta}
 \left(C_{VLL}^{K1\ell} +C_\text{SM}^{K1\ell}\right) (\bar{d}_L \ga^\mu s_L)(\bar{\nu}_{\ell L} \ga_\mu \nu_{\ell L}) +\hc, 
\end{align}
where
\begin{align}
 C_\text{SM}^{K1\ell} =& -C_F \left[\la_c X_{NL}^{\ell} +\la_t X(x_t)\right] ,\\
 C_{VLL}^{K1\ell} =& \frac{1}{2m_{S1}^2} \hat{Y}_1^L \hat{Y}_2^{L\ast} =\frac{\la}{2m_{S1}^2} \left( 1-\frac{\la^2}{2} \right) \left| \widetilde{Y}^L_2\right|^2 \de_{\ell\mu},
\end{align}
$\lam_{c(t)}=V_{c(t)s} V_{c(t)d}^{\ast}$\,, 
$X(x_t)$ (with $x_t=m^2_t/m_W^2$) is the top quark loop contribution to the effective operator, 
$X_{NL}^{\ell}$ is the charm-lepton $\ell$ loop contribution,
and
\begin{align}
 C_F = \frac{4G_F}{\sqrt{2}}\frac{\al}{2\pi s_W^2}\,.
\end{align}
Using the Wolfenstein parametrization one writes
\begin{align}
 \la_c =& -\la\left(1-\frac{\la^2}{2}\right),\\
 \la_t =& -A^2\la^5(1-\rh+i\eta).
\end{align}
The charm contributions are numerically computed, e.g, in Refs.\cite{Buchalla:1995vs,Brod:2008ss}. 
We use
\begin{align}
 X_{NL}^e =X_{NL}^\mu =10.6\times 10^{-4}\,,\qquad X_{NL}^\ta =7.01\times 10^{-4}\,,
\end{align}
and the top loop function is 
\begin{align}
 X(x) =\frac{x}{8} \left( \frac{x+2}{x-1} +\frac{3(x-2)}{(1-x)^2}\ln x \right).
\end{align}

Finally, we write down the branching ratio:
\begin{multline}
 \textrm{Br}(K^+\to\pi^+\nu\bar{\nu}) = \frac{\ka_+}{\la^{10}} \left\{ 
   \Im[\la_t]^2 X(x_t)^2
	+\frac{2}{3} \left(\Re[\la_t] X(x_t)+\la_c X_{NL}^e   \right)^2  \right. \\
\left.	+\frac{1}{3} \left(\Re[\la_t] X(x_t)+\la_c X_{NL}^\ta \right)^2
 +\frac{C_{VLL}^{K1\mu}}{3C_F}\left[ \frac{C_{VLL}^{K1\mu}}{C_F} -2\left(\Re[\la_t] X(x_t)+\la_c X_{NL}^e\right) \right]
 \right\},\label{brK2Pi}
\end{multline}
where $\ka_+=5.27\times 10^{-11}$\cite{0902.0160}.
The first three terms in \refeq{brK2Pi} yield the SM contribution, $\textrm{Br}(K^+\to\pi^+\nu\bar{\nu})_{\textrm{SM}} \approx 9 \times 10^{-11}$.
We get the $2\,\sigma$ bound
\begin{align}
 \left| \widetilde{Y}^L_2 \right| < 5.26 \times 10^{-2}\left(\frac{m_{S1}}{\tev}\right)\,.\label{K2Pibound}
\end{align}
The parameters used to obtain the numerical constraints are shown in Tab.~\ref{TabK}.

\begin{table}
\centering
\begin{tabular}{|c|c|}
\hline 
 $G_F$ & 1.17$\times 10^{-5} \gev^{-2}$  \\
 $\al$ & 1/137  \\
 $s_W^2$ & 0.231  \\
   $A$ & 0.836 \\
 $\rh$ & 0.135 \\
 $\eta$ & 0.349 \\
 $m_t$ &  173\gev  \\
 $m_W$ & 80.4\gev  \\
\hline
\end{tabular}
\qquad
\begin{tabular}{|c|c|}
\hline 
 $\ta_K$ & 5.12 $\times 10^{-8}$ sec. \\
   $m_K$ & 498\MeV \\
   $f_K$ & 156\MeV \\
   $m_s$ &  95\MeV \\
 \hline
\end{tabular}
\caption{
 The parameters used to obtain the experimental constraints from $\textrm{Br}(K^+\to\pi^+\nu\bar{\nu})$ and $\textrm{Br}(K_L\to\mu^+\mu^-)$.
 $G_F$, $\al$ and $s_W^2$ are given by CODATA\cite{Mohr:2015ccw}.
 For the other constants, the PDG values are employed\cite{Tanabashi:2018oca}.
}
\label{TabK}
\end{table}

As we shall better see in \refsec{sec:results}, the bound of \refeq{K2Pibound} implies that large $R$-type couplings are required in Model~1  
to be consistent with the BNL 
value for \deltagmtwomu. Thus, the parameter space can be directly probed by the LHC.\smallskip
 
In Model~2, the induced Yukawa coupling is restricted by the measurement of $\textrm{Br}(K_L\to\mu^+\mu^-)$.
This process is dominated by the long-distance contribution through $K_L\to\ga\ga$.
The 90\%~C.L. upper bound on the short distance contribution is given in Ref.\cite{Isidori:2003ts} 
as $\text{Br}(K_L\to\mu^+\mu^-)_\text{SD}<2.5\times 10^{-9}$.

The process is generated by the following effective interaction\cite{Kumar:2016omp}:
\begin{align}
 \mcl{L}_{K\mu\mu} = \left(C_{VLA}^{K2} +C_\text{SM}^{K2}\right) (\bar{d}_L \ga^\mu s_L)(\bar{\mu} \ga_\mu\ga_5 \mu) +\hc,
\end{align}
where
\begin{align}
 C_\text{SM}^{K2} =& -\frac{C_F}{2} \left[\la_c Y_{NL} +\la_t Y(x_t)\right],\\
 C_{VLA}^{K2} =& \frac{1}{4m_{R_2}^2} \hat{Y}_1^R \hat{Y}_2^{R\ast} = \frac{\la}{4m_{R_2}^2} \left( 1-\frac{\la^2}{2} \right) \left| \widetilde{Y}_2^R\right|^2,
\end{align}
with $Y_{NL}$ representing the contribution of loop diagrams involving the charm, and
$Y(x_t)$ the contribution from top quark loops. 
The charm contribution is $Y_{NL}=3.50\times 10^{-4}$\cite{Buchalla:1995vs}, and the top loop function is
\begin{align}
 Y (x) = \frac{x}{8} \left( \frac{x-4}{x-1} +\frac{3x}{(1-x)^2}\ln x \right).
\end{align}

Neglecting the tiny $CP$-violating contribution, we write down the short-distance branching ratio:
\be
 \text{Br}(K_L\to\mu^+\mu^-)_\text{SD} =
 \ta_K \frac{f_K^2 m_K^3}{4\pi}\frac{m_\mu^2}{m_K^2}\sqrt{1-\frac{4m_\mu^2}{m_K^2}} \Re\left(C_{VLA}^{K2}+2C_\text{SM}^{K2}\right)^2,
\ee
where lifetime $\tau_K$ and other constants relevant for the calculation are given in \reftable{TabK}.  
The SM contribution is $ \text{Br}(K_L\to\mu^+\mu^-)_\text{SM} =1.08\times 10^{-9}$.
As a result, we obtain the following constraint:
\begin{align}
 \left| \widetilde{Y}_2^R\right| < 1.9\times 10^{-2}\left(\frac{m_{R_2}}{\tev}\right).\label{KLbound}
\end{align}\bigskip

We conclude this section with a note on EW precision bounds. Precision tests of the EW theory are not very constraining for charmphilic leptoquark scenarios. 
The presence of leptoquarks can modify the coupling $Z\to f\bar{f}$ and induce effects that can be picked up in data from the $Z$ line shape and 
asymmetry observables.
The analytical form of the corresponding loop contribution can be found, for example, in\cite{Bansal:2018nwp}.
Since, in the charmphilic assumption, the leptoquark couples only to the second generation of fermions, a loop correction to the effective $Z\to f\bar{f}$ 
coupling scales proportionally to the $Z$ boson threshold, $\sim |Y_2^{R,L}|^2m_Z^2/m_{S_1,R_2}^2$. Therefore, it 
does not introduce additional constraints on the parameter space 
allowed by \gmtwo\ and other flavor observables.

\section{LHC constraints on the parameter space\label{sec:results}}

The parameter space surviving flavor exclusion 
bounds in charmphilic leptoquark scenarios for \gmtwo\ falls squarely inside the reach of LHC searches.
At small-to-moderate Yukawa couplings, $Y\sim\mathcal{O}(0.1)$, 
leptoquarks are predominantly pair-produced with a cross section directly proportional to the QCD strong coupling. 
Therefore, in this case the collider exclusion bounds depend solely on the mass of the leptoquark $m_{\textrm{LQ}}$
and read $m_{\textrm{LQ}}\gtrsim 1.5 \tev$\cite{Aaboud:2016qeg,Diaz:2017lit,Sirunyan:2018ryt}.

For larger Yukawa couplings, $Y\sim\mathcal{O}(1)$, dilepton production with a $t$-channel leptoquark exchange, as well as single leptoquark production, 
can be directly probed over vast mass ranges, and provide complementary constraints\cite{Faroughy:2016osc,Greljo:2017vvb}. 
Very recently it was also pointed out\cite{Bansal:2018eha} that, for leptoquarks coupling to the first quark generation in particular, 
monolepton searches can provide bounds equivalent to those from dilepton. 
In the large mass regime, where QCD direct production saturates, dilepton signatures are particularly effective in constraining the second-generation couplings.
Generation-dependent constraints were recently provided in Refs.\cite{Raj:2016aky,Angelescu:2018tyl,Schmaltz:2018nls} 
by recasting the results of two 13\tev\ LHC searches for high-mass resonances in dilepton final state by the ATLAS Collaboration\cite{Aaboud:2017buh} and 
the CMS Collaboration\cite{Sirunyan:2018exx} for a large set of leptoquark models.

We perform an analysis along similar lines in this study, and apply it to the parameter space of the 
$S_1$ and $R_2$ models, allowed after incorporating the bounds from the measurement of \gmtwo\ and other flavor observables. 
Recall, in particular, that while we have shown that the down-origin case for \deltagmtwomu\ is 
excluded by the measurement of $\textrm{Br}(D^0\rightarrow \mu^+\mu^-)$, when it comes to the up-origin assumption 
Eqs.~(\ref{K2Pibound}) and (\ref{KLbound}) only bound one of the couplings entering the product $\Re(g_L g_R^{\ast})$ in \refeq{chiralgm2}.
The remaining part corresponds to relatively large values of the Yukawa couplings $Y_2^R$, $Y_2^L$, and hence it is subject to constraints from the LHC dimuon searches.

To perform the analysis, each leptoquark model was generated with \texttt{FeynRules}\cite{Alloul:2013bka} 
and the corresponding {\tt UFO} files were passed to \mad\cite{Alwall:2014hca}. The results were cross-checked with 
the code used in\cite{Kowalska:2017iqv}, based on \spheno\cite{Porod:2003um,Porod:2011nf}, \texttt{PYTHIA}\cite{Sjostrand:2007gs}, 
and \texttt{DELPHES~3}\cite{deFavereau:2013fsa}.
We generated two process, $pp\rightarrow \mu^+\mu^-$ and $p\,p\to \mu^+\mu^- j$, particularly the $g\,\overset{(-)}{c}\rightarrow \mu^+\mu^- \overset{(-)}{c}$ contribution (without the $c$-tag), 
which can produce a signal in dimuon searches, because jets in the generated events are not vetoed. 
Note that the inclusion of $p\,p\to \mu^+\mu^- j$ is particularly important, as in this case the invariant mass distribution acquires a long high-$p_T$ tail due to the 
gluon PDF. 
This effect cannot be observed in the pure $p\,p\to \mu^+\mu^-$ production.

\begin{figure}[t]
\centering
\subfloat[]{%
\includegraphics[width=0.47\textwidth]{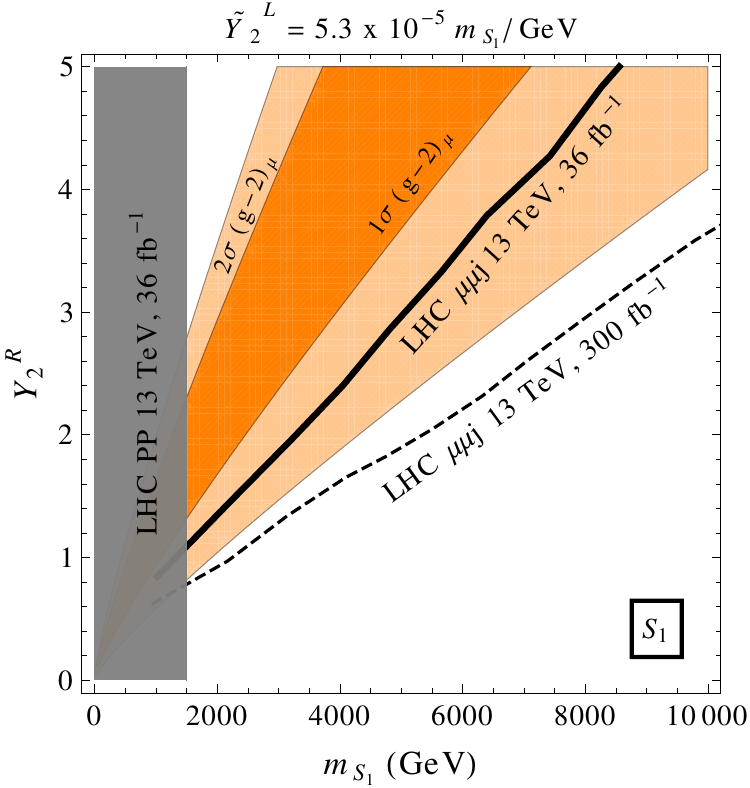}
}%
\subfloat[]{%
\includegraphics[width=0.47\textwidth]{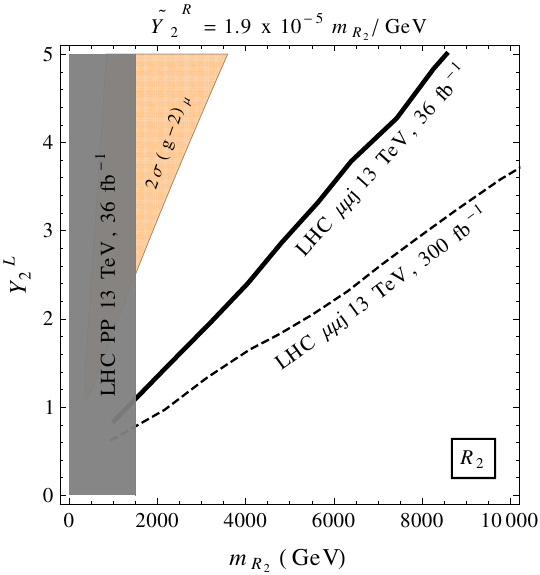}
}%
\caption{(a) Parameter space of the leptoquark model $S_1$ in the 
plane ($m_{S_1}$, $Y_2^R$) allowed by the measurement of \deltagmtwomu\ and the experimental bounds from $K^+\to\pi^+\nu\bar{\nu}$. The $1\,\sigma$
($2\,\sigma$) region is shown in orange (light orange). Solid black line indicates the 95\% \cl\ upper limit based on our recast of the CMS dilepton search at $36\invfb$\cite{Sirunyan:2018exx}, 
while the dashed black line indicates the projected sensitivity for $300\invfb$. 
Gray region represents the 95\%~\cl\ exclusion bound due to the leptoquark pair-production\cite{Aaboud:2016qeg,Diaz:2017lit,Sirunyan:2018ryt}. 
(b) Same for the parameter space of the leptoquark model $R_2$ in the plane ($m_{R_2}$, $Y_2^L$) allowed by the measurement of \deltagmtwomu\ 
and the experimental bounds from the short-distance contribution to $K_L \rightarrow \mu^+\mu^-$.}
\label{fig:LHCbounds}
\end{figure}

Six kinematical bins based on the invariant dimuon mass, $m_{\mu\mu}$, were constructed, closely following the CMS search for high-mass resonances in dilepton 
final states\cite{Sirunyan:2018exx}. For each point in the parameter space we calculate the likelihood function, using a Poisson distribution smeared with the experimental background determination uncertainty provided in Ref.\cite{Sirunyan:2018exx}, and statistically combining the six exclusive kinematical bins of the invariant dimuon mass by multiplying the individual likelihood functions. 
The 95\%~\cl\ exclusion limit was derived using the $\Delta\chisq$ 
statistics. The dominant bins affecting the obtained bound were the three highest by invariant mass, $m_{\mu\mu}>900\gev$.
 
We show in \reffig{fig:LHCbounds}(a) the 95\%~\cl\ upper bound from the LHC in solid black.\footnote{Our 36\invfb\ bound is in good agreement with the recent estimate 
of Ref.\cite{Angelescu:2018tyl} after we include the process $pp\rightarrow \mu^+\mu^- j$ in our simulation.
If one only includes $p p(c\bar{c})\rightarrow \mu^+\mu^-$ the bound weakens,
and it agrees with the recent computation of Ref.\cite{Schmaltz:2018nls}.} 
Gray region represents the present 95\%~\cl\ exclusion due to leptoquark pair-production\cite{Aaboud:2016qeg,Diaz:2017lit,Sirunyan:2018ryt}.
In dark orange we show the $1\,\sigma$ region for \deltagmtwomu\ in Model~1 after the strongest flavor constraint, \refeq{K2Pibound}, has been taken into account.
We show in light orange the corresponding $2\,\sigma$ region.
The dashed black line 
indicates our projected LHC sensitivity with $300\invfb$, which entirely excludes the $2\,\sigma$ region.
Note, moreover, that the currently running NA62 experiment is expected to improve the
$\textrm{Br}(K^+\to\pi^+\nu\bar{\nu})$ measurement down to a precision of about 10\%\cite{Anelli:2005ju} 
and thus further tighten the bounds on this scenario.

In \reffig{fig:LHCbounds}(b) we compare our 95\%~\cl\ upper bound (black solid) with the allowed $2\,\sigma$ parameter space (light orange) in Model~2,
after the constraint from \refeq{KLbound} has been taken into account. The model is already excluded by the LHC. 
In order to explain the \gmtwo\ anomaly and at the same time avoid the strong 
tension with this bound one would
need an additional coupling to the top quark of the size of the coupling to the charm. 
Accidentally, the minimally required $\Re(\widetilde{Y}^R_3 Y_3^{L\ast})$ in this case 
coincides with the amount required to evade the $D^0\to \mu^+\mu^-$ bound in the down-origin scenario, \refeq{topphil}. 

\begin{figure}[t]
\centering
\includegraphics[width=0.70\textwidth]{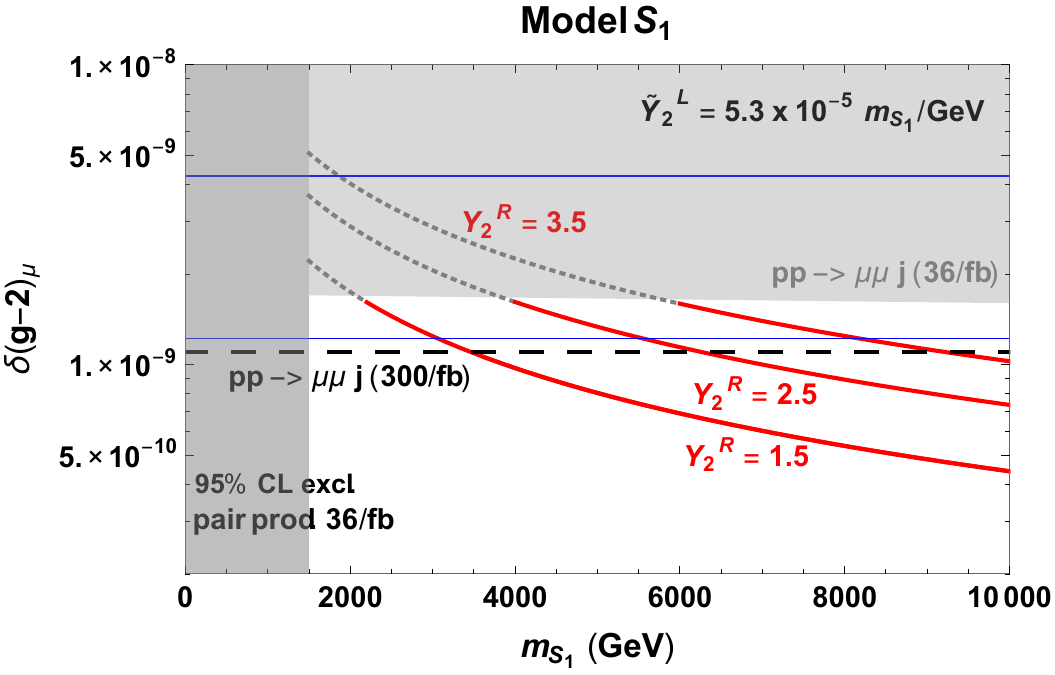}
\caption{The value of \deltagmtwomu\ after including the bounds from $\textrm{Br}(K^+\rightarrow \pi^+\nu\bar{\nu})$, as a function of leptoquark mass $m_{S_1}$, confronted with 
existing and projected LHC bounds. The solid blue lines delimit the $2\,\sigma$ region of the BNL experiment. 
The dark gray region shows the lower mass bound from Refs.\cite{Aaboud:2016qeg,Diaz:2017lit,Sirunyan:2018ryt}. In light gray, we present our recasting of the 
dilepton bound from Ref.\cite{Sirunyan:2018exx}, applied to $g\,\overset{(-)}{c}\rightarrow \mu^+\mu^- \overset{(-)}{c}$ processes. Our projection for 300\invfb\ is shown as a black dashed line.}
\label{fig:g2_mod1}
\end{figure}

Finally we summarize the overall collider situation 
in \reffig{fig:g2_mod1}, which shows a plot of the deviation from the SM, \deltagmtwomu, as a function of leptoquark mass in Model~1, confronted with our calculated LHC bounds and projections.
One can see that the reach of the LHC searches pushes down along the $y$-axis as the luminosity increases. Thus, 
the LHC will continue to provide, with its High-Luminosity run, the most effective tool to probe these scenarios 
for a wide range of possible outcomes at Fermilab. 

\section{Summary and conclusions\label{sec:summary}}

We have considered in this work charmphilic contributions to the muon anomalous magnetic moment with a single leptoquark. 
As the only way to increase the value of \gmtwo, and at the same time keep the leptoquark mass beyond direct-production bounds at colliders,
is in this case to couple to the quark with a log-enhanced interaction, 
only two models are relevant for this observable: the scalar leptoquarks $S_1$ and $R_2$. 
We have confronted these two models with relevant constraints from flavor measurements and numerically 
recast the LHC searches that can provide applicable limits. 

In order to systematically address the flavor bounds, 
we have subdivided the available parameter space of these models into
two regions: ``up-origin,'' if the charmphilic ansatz applies to the leptoquark coupling to the muon and to the up-type component of the second generation quark doublet, and the ``down-origin," 
if it applies to the coupling with the muon and the down-type component quark. 

We find that, under the down-origin assumption,
the parameter space consistent with the measured \gmtwo\ anomaly at BNL is in both models entirely excluded by the measurement of $\textrm{Br}(D^0\rightarrow \mu^+\mu^-)$, 
unless one introduces a coupling to the top quark of the same order of magnitude as the charm's.    

Conversely, under the up-origin assumption some of the parameter space of the model with $R_2$ survives the most constraining flavor bound, from the short-distance contribution to
$\textrm{Br}(K_L\rightarrow \mu^+\mu^-)$, but is by now entirely excluded by searches with two muons in the final state at the LHC. The bound can be evaded by assuming 
the existence of couplings of both the chiral states of the muon to the top quark,
of approximately the same size as the couplings to the charm. 
 
Finally, the leptoquark $S_1$ is the least constrained under the up-origin assumption. 
Some of the parameter space survives the most constraining flavor bound, from $\textrm{Br}(K^+\rightarrow \pi^+\nu\bar{\nu})$, and also some of the BNL $2\,\sigma$ region remains in play 
after the LHC dimuon searches are taken into account. However, once $300\invfb$ of luminosity are accumulated, 
the remaining parameter space will be probed in its entirety.  

\bigskip
\noindent \textbf{ACKNOWLEDGMENTS}
\medskip

\noindent We would like to thank K.~Sakurai for his comments on the manuscript. 
YY would like to thank M. Tanaka and K. Hamaguchi for directing his interest to leptoquarks and the muon g--2.
KK is supported in part by the National Science Centre (Poland) under the research Grant No.~2017/26/E/ST2/00470.
EMS and YY are supported in part by the National Science Centre (Poland) under the research Grant No.~2017/26/D/ST2/00490. 
The use of the CIS computer cluster at the National Centre for Nuclear Research in Warsaw is gratefully acknowledged.

\bigskip
\newpage
\appendix

\addcontentsline{toc}{section}{Appendices}

\section{Update after the new $\mathbf{\gmtwo}$ measurement \label{app:update}}

The anomalous magnetic moment of the muon was recently measured by the E989 experiment at Fermilab\cite{PhysRevLett.126.141801}. The collaboration reports
\be
a_{\mu}^{\textrm{E989}}=116592040(54)\times 10^{-11}\label{eq:Fer}\,,
\ee
indicating a $3.3\,\sigma$ deviation from the value expected in the Standard Model~(SM)\cite{Davier:2017zfy,Keshavarzi:2018mgv,Colangelo:2018mtw,Hoferichter:2019mqg,Davier:2019can,Keshavarzi:2019abf,Kurz:2014wya,Melnikov:2003xd,Masjuan:2017tvw,Colangelo:2017fiz,Hoferichter:2018kwz,Gerardin:2019vio,Bijnens:2019ghy,Colangelo:2019uex,Colangelo:2014qya,Blum:2019ugy,Aoyama:2012wk,atoms7010028,Czarnecki:2002nt,Gnendiger:2013pva}, 
\be
a_{\mu}^{\textrm{SM}}=116591810(43)\times 10^{-11}\label{eq:SM}\,.
\ee

The result has been long awaited because the previous experimental determination,
obtained a couple of decades ago at Brookhaven National Lab (BNL)\cite{Bennett:2006fi}, 
\be
a_{\mu}^{\textrm{BNL}}=116592089(63)\times 10^{-11}\label{eq:BNL}\,,
\ee
also showed a $\sim 3.7\sigma$ discrepancy.

When Eqs.~(\ref{eq:Fer}) and (\ref{eq:BNL}) are statistically combined 
one currently obtains\cite{PhysRevLett.126.141801}
\be\label{eq:meas}
\deltagmtwomu=\left(2.51\pm 0.59 \right)\times 10^{-9}\,,
\ee
which yields a deviation from the SM at the $4.2\,\sigma$ level. 

 In light of the new \gmtwo\ measurement from the Fermilab experiment~(\ref{eq:Fer}), and taking into account the relatively recent release of LHC analyses with $\sim 140\,\textrm{fb}^{-1}$ luminosity, in this appendix we update our previous results with a new computation of the relevant parameter space regions and a new recasting of the 
$140\,\textrm{fb}^{-1}$ luminosity data at the LHC.

\paragraph{Model $\boldsymbol{S_1}$} Given the value in \refeq{eq:meas} we get the $2\sigma$ bound
\be
3.4\times 10^{-2}\left(\frac{m_{S_1}}{\tev}\right)^2\leq \Re{(\widetilde{Y}^L_2 Y_2^{R\ast})}\left(1+0.17 \ln\frac{m_{S_1}}{\tev}\right)\leq 9.3\times 10^{-2}\left(\frac{m_{S_1}}{\tev}\right)^2.\label{g2boundMod1new}
\ee

By comparing \refeq{D0bound} with \refeq{g2boundMod1new} one can see 
that the $D_0\to \mu^+ \mu^-$ constraint still excludes entirely 
a charmphilic solution the \gmtwo\ anomaly, even after the new determination at Fermilab. In order to avoid the bound while still inducing the appropriate value for the anomalous magnetic moment one is forced to introduce 
a small coupling to the top quark. 

Equation~(\ref{g2boundMod1new}) then becomes
\be
3.4\times 10^{-2}\left(\frac{m_{S_1}}{\tev}\right)^2\leq a_c \Re{(\widetilde{Y}^L_2 Y_2^{R\ast})}+20.7\, a_t \Re{(\widetilde{Y}^L_3 Y_3^{R\ast})}\leq 9.3\times 10^{-2}\left(\frac{m_{S_1}}{\tev}\right)^2,\label{eq:top_S1}
\ee
where $a_c=1+0.17\ln(m_{S_1}/\textrm{TeV})$ and $a_t=1+1.06\ln(m_{S_1}/\textrm{TeV})$. 

We get
the minimum topphilic requirement for $S_1$ given \refeq{D0bound}:
\be
\textrm{Re}(\widetilde{Y}^L_3 Y^{R\ast}_3)\gsim 1.1\times 10^{-3}\left(\frac{m_{S_1}}{\textrm{TeV}}\right)^2\times\frac{1}{a_t}\left[1-0.08 \ln\left(\frac{m_{S_1}}{\textrm{TeV}}\right)\right].
\ee

\paragraph{Correction to \refeq{K2Pibound}} We correct a sign error in the analysis leading to \refeq{K2Pibound}. The correct sign gives
\be
\hat{Y}^L_1\hat{Y}^{L\ast}_2= -\lam\left(1-\frac{\lam^2}{2}\right)\left|\widetilde{Y}^L_2\right|^2\,.
\ee

The bound is modified as
\be\label{K2Piboundnew}
\left|\widetilde{Y}^L_2 \right|\lesssim 7.1\times 10^{-2}\left(\frac{m_{S_1}}{\textrm{TeV}} \right).
\ee

\paragraph{Model $\boldsymbol{R_2}$} Given the value in \refeq{eq:meas} we get the $2\sigma$ bound
\be\label{g2boundMod2new}
3.0\times 10^{-2}\left(\frac{m_{R_2}}{\tev}\right)^2\leq -\Re{(Y^L_2 \widetilde{Y}_2^{R\ast})}\left(1+0.15 \ln\frac{m_{R_2}}{\tev}\right)\leq 8.2\times 10^{-2}\left(\frac{m_{R_2}}{\tev}\right)^2.
\ee

We can calculate again the minimum amount of topphilic content required to avoid the bound in \refeq{D0bound}. Equation~(\ref{g2boundMod2}) becomes
\be\label{eq:top_R2}
3.0\times 10^{-2}\left(\frac{m_{R_2}}{\tev}\right)^2\leq -a_c\Re{(Y^L_2 \widetilde{Y}_2^{R\ast})}-33.0\,a_t\Re{(Y_3^L \widetilde{Y}_3^{R\ast})} \leq 8.2\times 10^{-2}\left(\frac{m_{R_2}}{\tev}\right)^2,
\ee
where $a_c=1+0.15\ln(m_{R_2}/\textrm{TeV})$ and $a_t=1+0.59\ln(m_{R_2}/\textrm{TeV})$. 

We get
the minimum topphilic requirement for $R_2$:
\be
-\textrm{Re}(\widetilde{Y}^{R\ast}_3 Y^L_3)\gsim 5.7\times 10^{-4}\left(\frac{m_{R_2}}{\textrm{TeV}}\right)^2\times\frac{1}{a_t}\left[1-0.09 \ln\left(\frac{m_{R_2}}{\textrm{TeV}}\right)\right]\,.
\ee

\subsection{LHC constraints \label{app:LHC}}

We closely follow the approach described in detail in the previous analysis. We have implemented each LQ 
model with \texttt{FeynRules}
and we have passed the corresponding {\tt UFO} files to \mad, generating the two processes contributing to the signal, $pp\rightarrow \mu^+\mu^-$ and $p\,p\to \mu^+\mu^- j$. 

The CMS Collaboration has recently updated the search for high-mass resonances in the di-lepton final state with $139\invfb$\cite{Sirunyan:2021khd}, thus providing a
factor-of-four increase in integrated luminosity with respect to their previous result\cite{Sirunyan:2018exx}. We derived the 95\%~\cl\ upper bound in the plane of the LQ Yukawa coupling versus mass, which was meant to supersede the corresponding limit 
presented in \reffig{fig:LHCbounds}. 
As the current CMS bound does not change significantly, however, 
we do not report it over here. 
The reason for this weak improvement in exclusion power lies on the strong,   
greater than~$2\sigma$ downward fluctuations in the number of observed events that appeared in the two highest di-muon mass bins of the $36\invfb$ data set\cite{Sirunyan:2018exx} and were not confirmed in Ref.\cite{Sirunyan:2021khd}. 

For additional constraining power we thus recast the search for high-mass di-lepton resonances with $139\invfb$ of data at ATLAS\cite{Aad:2019fac}.
The experimental collaboration provides the analytical form of the background as a function of the di-muon invariant mass, as well as the binned observed 
di-muon invariant mass spectrum.
In order to construct the likelihood function in a counting-experiment fashion, we 
have transformed the functional form of the background into a binned background distribution normalized in the same way as the observed di-muon mass distribution. 
Seven kinematical bins were constructed, closely following the previous ATLAS search for high-mass resonances in di-lepton final states\cite{Aaboud:2017buh}. 
For each point in the Yukawa-coupling vs.~mass parameter space we have calculated the likelihood function using a Poisson distribution, and statistically combining the invariant mass bins. 

\begin{figure}[t]
\centering
\subfloat[]{%
\includegraphics[width=0.47\textwidth]{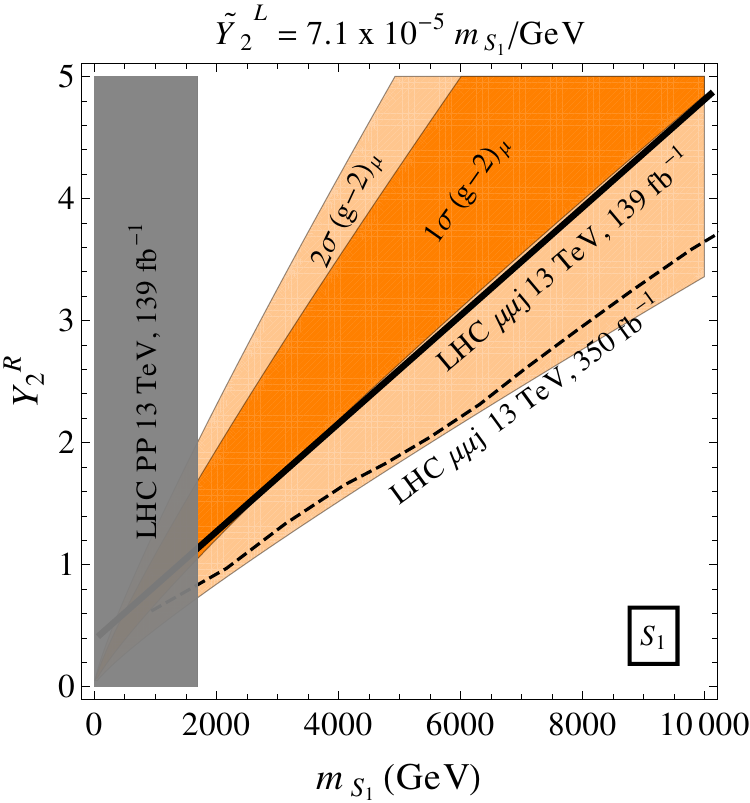}
}%
\subfloat[]{%
\includegraphics[width=0.47\textwidth]{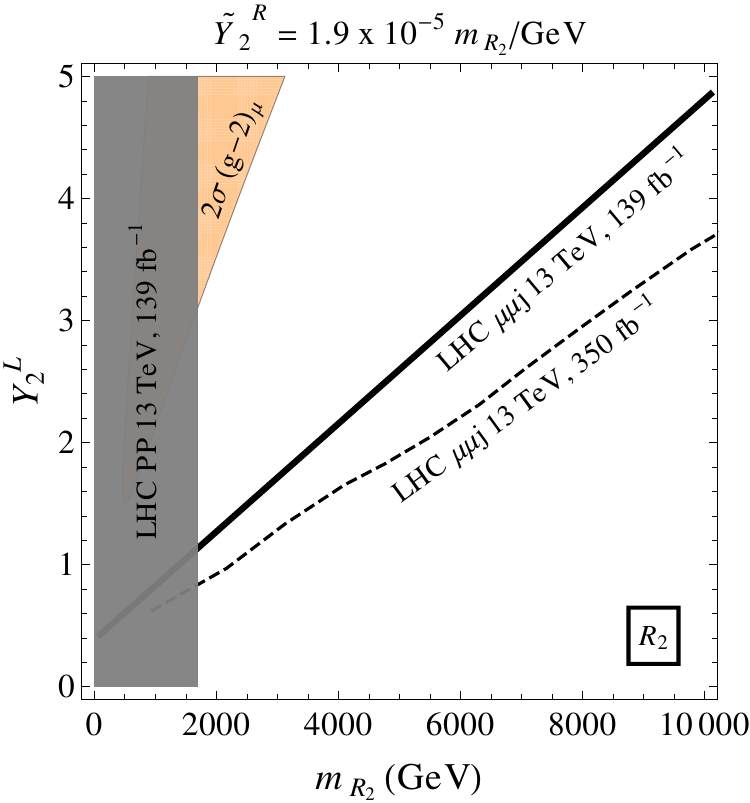}
}%
\caption{(a) Experimental status of the leptoquark model $S_1$ in the 
plane ($m_{S_1}$, $Y_2^R$). The $1\,\sigma$
($2\,\sigma$) region allowed by the statistical combination of the measurements of \deltagmtwomu\ and the experimental bounds from $K^+\to\pi^+\nu\bar{\nu}$ is marked in orange (light orange). Solid black line shows the 95\% \cl\ upper limit based on our recast of the ATLAS di-lepton search at $139\invfb$\cite{Aad:2019fac}, 
while the projected sensitivity for $350\invfb$ is indicated as a dashed black line. 
Gray vertical band shows the 95\%~\cl\ exclusion bound given by the leptoquark pair-production search by ATLAS at $139\invfb$\cite{Aad:2020iuy}. 
(b) Experimental status of the leptoquark model $R_2$ in the plane ($m_{R_2}$, $Y_2^L$) after inclusion of the experimental bounds from the measurements of \deltagmtwomu\ 
and  the short-distance contribution to $K_L \rightarrow \mu^+\mu^-$.}
\label{fig:LHCboundsnew}
\end{figure}

In \reffig{fig:LHCboundsnew}(a) we present the updated experimental status 
in the $S_1$ case. In solid black the 95\%~\cl\ upper bound from the ATLAS di-lepton search\cite{Aad:2019fac} is indicated, while the dashed black line shows our projected LHC sensitivity with $350\invfb$. Gray vertical band corresponds to the most recent  95\%~\cl\ exclusion due to the leptoquark pair-production search by ATLAS\cite{Aad:2020iuy}.  The part of parameter space favored at $1\,\sigma$ by the combination of \deltagmtwomu\ measurements after the strongest flavor constraint, \refeq{K2Piboundnew}, has been taken into account, is shown in dark orange. The corresponding $2\,\sigma$ region is marked in light orange. More than $350\invfb$ of the LHC data will be needed to entirely test this part of the parameter space. 

The corresponding summary in the case of $R_2$ is shown in \reffig{fig:LHCboundsnew}(b), after the constraint from \refeq{KLbound} has been taken into account. This scenario remains strongly excluded by the LHC searches even after the new \gmtwo\ measurement.

\bibliographystyle{utphysmcite}
\bibliography{KSY1}

\ifx\mcitethebibliography\mciteundefinedmacro
\PackageError{unsrtM.bst}{mciteplus.sty has not been loaded}
{This bibstyle requires the use of the mciteplus package.}\fi
\begin{mcitethebibliography}{100}

\bibitem{Grange:2015fou}
{\bfseries Muon g-2} Collaboration, J.~Grange {\em et~al.}, ``{Muon (g-2)
  Technical Design Report},''
\href{http://arxiv.org/abs/1501.06858}{{\ttfamily arXiv:1501.06858
  [physics.ins-det]}}.
\mciteBstWouldAddEndPunctfalse
\mciteSetBstMidEndSepPunct{\mcitedefaultmidpunct}
{}{\mcitedefaultseppunct}\relax
\EndOfBibitem
\bibitem{Mibe:2010zz}
{\bfseries J-PARC g-2} Collaboration, T.~Mibe, ``{New g-2 experiment at
  J-PARC},''
\href{http://dx.doi.org/10.1088/1674-1137/34/6/022}{{\em Chin. Phys.}
  {\bfseries C34} (2010) 745--748}.
\mciteBstWouldAddEndPunctfalse
\mciteSetBstMidEndSepPunct{\mcitedefaultmidpunct}
{}{\mcitedefaultseppunct}\relax
\EndOfBibitem
\bibitem{Bennett:2006fi}
{\bfseries Muon g-2} Collaboration, G.~W. Bennett {\em et~al.}, ``{Final Report
  of the Muon E821 Anomalous Magnetic Moment Measurement at BNL},''
  \href{http://dx.doi.org/10.1103/PhysRevD.73.072003}{{\em Phys. Rev.}
  {\bfseries D73} (2006) 072003},
\href{http://arxiv.org/abs/hep-ex/0602035}{{\ttfamily arXiv:hep-ex/0602035
  [hep-ex]}}.
\mciteBstWouldAddEndPunctfalse
\mciteSetBstMidEndSepPunct{\mcitedefaultmidpunct}
{}{\mcitedefaultseppunct}\relax
\EndOfBibitem
\bibitem{Davier:2016iru}
M.~Davier, ``{Update of the Hadronic Vacuum Polarisation Contribution to the
  muon g-2},'' \href{http://dx.doi.org/10.1016/j.nuclphysbps.2017.03.047}{{\em
  Nucl. Part. Phys. Proc.} {\bfseries 287-288} (2017) 70--75},
\href{http://arxiv.org/abs/1612.02743}{{\ttfamily arXiv:1612.02743 [hep-ph]}}.
\mciteBstWouldAddEndPunctfalse
\mciteSetBstMidEndSepPunct{\mcitedefaultmidpunct}
{}{\mcitedefaultseppunct}\relax
\EndOfBibitem
\bibitem{Jegerlehner:2017lbd}
F.~Jegerlehner, ``{Muon g-2 theory: The hadronic part},''
  \href{http://dx.doi.org/10.1051/epjconf/201816600022}{{\em EPJ Web Conf.}
  {\bfseries 166} (2018) 00022},
\href{http://arxiv.org/abs/1705.00263}{{\ttfamily arXiv:1705.00263 [hep-ph]}}.
\mciteBstWouldAddEndPunctfalse
\mciteSetBstMidEndSepPunct{\mcitedefaultmidpunct}
{}{\mcitedefaultseppunct}\relax
\EndOfBibitem
\bibitem{Moroi:1995yh}
T.~Moroi, ``{The Muon anomalous magnetic dipole moment in the minimal
  supersymmetric standard model},''
  \href{http://dx.doi.org/10.1103/PhysRevD.53.6565,
  10.1103/PhysRevD.56.4424}{{\em Phys. Rev.} {\bfseries D53} (1996)
  6565--6575}, \href{http://arxiv.org/abs/hep-ph/9512396}{{\ttfamily
  arXiv:hep-ph/9512396 [hep-ph]}}.
[Erratum: Phys. Rev.D56,4424(1997)].
\mciteBstWouldAddEndPunctfalse
\mciteSetBstMidEndSepPunct{\mcitedefaultmidpunct}
{}{\mcitedefaultseppunct}\relax
\EndOfBibitem
\bibitem{Cho:2000sf}
G.-C. Cho, K.~Hagiwara, and M.~Hayakawa, ``{Muon g-2 and precision electroweak
  physics in the MSSM},''
  \href{http://dx.doi.org/10.1016/S0370-2693(00)00288-4}{{\em Phys. Lett.}
  {\bfseries B478} (2000) 231--238},
\href{http://arxiv.org/abs/hep-ph/0001229}{{\ttfamily arXiv:hep-ph/0001229
  [hep-ph]}}.
\mciteBstWouldAddEndPunctfalse
\mciteSetBstMidEndSepPunct{\mcitedefaultmidpunct}
{}{\mcitedefaultseppunct}\relax
\EndOfBibitem
\bibitem{Martin:2001st}
S.~P. Martin and J.~D. Wells, ``{Muon Anomalous Magnetic Dipole Moment in
  Supersymmetric Theories},''
  \href{http://dx.doi.org/10.1103/PhysRevD.64.035003}{{\em Phys. Rev.}
  {\bfseries D64} (2001) 035003},
\href{http://arxiv.org/abs/hep-ph/0103067}{{\ttfamily arXiv:hep-ph/0103067
  [hep-ph]}}.
\mciteBstWouldAddEndPunctfalse
\mciteSetBstMidEndSepPunct{\mcitedefaultmidpunct}
{}{\mcitedefaultseppunct}\relax
\EndOfBibitem
\bibitem{Aaboud:2018jiw}
{\bfseries ATLAS} Collaboration, M.~Aaboud {\em et~al.}, ``{Search for
  electroweak production of supersymmetric particles in final states with two
  or three leptons at $\sqrt{s}=13\,$TeV with the ATLAS detector},''
  \href{http://dx.doi.org/10.1140/epjc/s10052-018-6423-7}{{\em Eur. Phys. J.}
  {\bfseries C78} no.~12, (2018) 995},
\href{http://arxiv.org/abs/1803.02762}{{\ttfamily arXiv:1803.02762 [hep-ex]}}.
\mciteBstWouldAddEndPunctfalse
\mciteSetBstMidEndSepPunct{\mcitedefaultmidpunct}
{}{\mcitedefaultseppunct}\relax
\EndOfBibitem
\bibitem{Aaboud:2017leg}
{\bfseries ATLAS} Collaboration, M.~Aaboud {\em et~al.}, ``{Search for
  electroweak production of supersymmetric states in scenarios with compressed
  mass spectra at $\sqrt{s}=13$ TeV with the ATLAS detector},''
  \href{http://dx.doi.org/10.1103/PhysRevD.97.052010}{{\em Phys. Rev.}
  {\bfseries D97} no.~5, (2018) 052010},
\href{http://arxiv.org/abs/1712.08119}{{\ttfamily arXiv:1712.08119 [hep-ex]}}.
\mciteBstWouldAddEndPunctfalse
\mciteSetBstMidEndSepPunct{\mcitedefaultmidpunct}
{}{\mcitedefaultseppunct}\relax
\EndOfBibitem
\bibitem{Sirunyan:2017lae}
{\bfseries CMS} Collaboration, A.~M. Sirunyan {\em et~al.}, ``{Search for
  electroweak production of charginos and neutralinos in multilepton final
  states in proton-proton collisions at $\sqrt{s}=$ 13 TeV},''
  \href{http://dx.doi.org/10.1007/JHEP03(2018)166}{{\em JHEP} {\bfseries 03}
  (2018) 166},
\href{http://arxiv.org/abs/1709.05406}{{\ttfamily arXiv:1709.05406 [hep-ex]}}.
\mciteBstWouldAddEndPunctfalse
\mciteSetBstMidEndSepPunct{\mcitedefaultmidpunct}
{}{\mcitedefaultseppunct}\relax
\EndOfBibitem
\bibitem{Sirunyan:2018iwl}
{\bfseries CMS} Collaboration, A.~M. Sirunyan {\em et~al.}, ``{Search for new
  physics in events with two soft oppositely charged leptons and missing
  transverse momentum in proton-proton collisions at $\sqrt{s}=$ 13 TeV},''
  \href{http://dx.doi.org/10.1016/j.physletb.2018.05.062}{{\em Phys. Lett.}
  {\bfseries B782} (2018) 440--467},
\href{http://arxiv.org/abs/1801.01846}{{\ttfamily arXiv:1801.01846 [hep-ex]}}.
\mciteBstWouldAddEndPunctfalse
\mciteSetBstMidEndSepPunct{\mcitedefaultmidpunct}
{}{\mcitedefaultseppunct}\relax
\EndOfBibitem
\bibitem{Endo:2013bba}
M.~Endo, K.~Hamaguchi, S.~Iwamoto, and T.~Yoshinaga, ``{Muon g-2 vs LHC in
  Supersymmetric Models},''
  \href{http://dx.doi.org/10.1007/JHEP01(2014)123}{{\em JHEP} {\bfseries 01}
  (2014) 123},
\href{http://arxiv.org/abs/1303.4256}{{\ttfamily arXiv:1303.4256 [hep-ph]}}.
\mciteBstWouldAddEndPunctfalse
\mciteSetBstMidEndSepPunct{\mcitedefaultmidpunct}
{}{\mcitedefaultseppunct}\relax
\EndOfBibitem
\bibitem{Akula:2013ioa}
S.~Akula and P.~Nath, ``{Gluino-driven radiative breaking, Higgs boson mass,
  muon g-2, and the Higgs diphoton decay in supergravity unification},''
  \href{http://dx.doi.org/10.1103/PhysRevD.87.115022}{{\em Phys. Rev.}
  {\bfseries D87} no.~11, (2013) 115022},
\href{http://arxiv.org/abs/1304.5526}{{\ttfamily arXiv:1304.5526 [hep-ph]}}.
\mciteBstWouldAddEndPunctfalse
\mciteSetBstMidEndSepPunct{\mcitedefaultmidpunct}
{}{\mcitedefaultseppunct}\relax
\EndOfBibitem
\bibitem{Fowlie:2013oua}
A.~Fowlie, K.~Kowalska, L.~Roszkowski, E.~M. Sessolo, and Y.-L.~S. Tsai,
  ``{Dark matter and collider signatures of the MSSM},''
  \href{http://dx.doi.org/10.1103/PhysRevD.88.055012}{{\em Phys. Rev.}
  {\bfseries D88} (2013) 055012},
\href{http://arxiv.org/abs/1306.1567}{{\ttfamily arXiv:1306.1567 [hep-ph]}}.
\mciteBstWouldAddEndPunctfalse
\mciteSetBstMidEndSepPunct{\mcitedefaultmidpunct}
{}{\mcitedefaultseppunct}\relax
\EndOfBibitem
\bibitem{Endo:2013lva}
M.~Endo, K.~Hamaguchi, T.~Kitahara, and T.~Yoshinaga, ``{Probing Bino
  contribution to muon $g - 2$},''
  \href{http://dx.doi.org/10.1007/JHEP11(2013)013}{{\em JHEP} {\bfseries 11}
  (2013) 013},
\href{http://arxiv.org/abs/1309.3065}{{\ttfamily arXiv:1309.3065 [hep-ph]}}.
\mciteBstWouldAddEndPunctfalse
\mciteSetBstMidEndSepPunct{\mcitedefaultmidpunct}
{}{\mcitedefaultseppunct}\relax
\EndOfBibitem
\bibitem{Chakraborti:2014gea}
M.~Chakraborti, U.~Chattopadhyay, A.~Choudhury, A.~Datta, and S.~Poddar, ``{The
  Electroweak Sector of the pMSSM in the Light of LHC - 8 TeV and Other
  Data},'' \href{http://dx.doi.org/10.1007/JHEP07(2014)019}{{\em JHEP}
  {\bfseries 07} (2014) 019},
\href{http://arxiv.org/abs/1404.4841}{{\ttfamily arXiv:1404.4841 [hep-ph]}}.
\mciteBstWouldAddEndPunctfalse
\mciteSetBstMidEndSepPunct{\mcitedefaultmidpunct}
{}{\mcitedefaultseppunct}\relax
\EndOfBibitem
\bibitem{Kowalska:2015zja}
K.~Kowalska, L.~Roszkowski, E.~M. Sessolo, and A.~J. Williams, ``{GUT-inspired
  SUSY and the muon g-2 anomaly: prospects for LHC 14 TeV},''
  \href{http://dx.doi.org/10.1007/JHEP06(2015)020}{{\em JHEP} {\bfseries 06}
  (2015) 020},
\href{http://arxiv.org/abs/1503.08219}{{\ttfamily arXiv:1503.08219 [hep-ph]}}.
\mciteBstWouldAddEndPunctfalse
\mciteSetBstMidEndSepPunct{\mcitedefaultmidpunct}
{}{\mcitedefaultseppunct}\relax
\EndOfBibitem
\bibitem{Padley:2015uma}
B.~P. Padley, K.~Sinha, and K.~Wang, ``{Natural Supersymmetry, Muon $g-2$, and
  the Last Crevices for the Top Squark},''
  \href{http://dx.doi.org/10.1103/PhysRevD.92.055025}{{\em Phys. Rev.}
  {\bfseries D92} no.~5, (2015) 055025},
\href{http://arxiv.org/abs/1505.05877}{{\ttfamily arXiv:1505.05877 [hep-ph]}}.
\mciteBstWouldAddEndPunctfalse
\mciteSetBstMidEndSepPunct{\mcitedefaultmidpunct}
{}{\mcitedefaultseppunct}\relax
\EndOfBibitem
\bibitem{Kowalska:2017iqv}
K.~Kowalska and E.~M. Sessolo, ``{Expectations for the muon g-2 in simplified
  models with dark matter},''
  \href{http://dx.doi.org/10.1007/JHEP09(2017)112}{{\em JHEP} {\bfseries 09}
  (2017) 112},
\href{http://arxiv.org/abs/1707.00753}{{\ttfamily arXiv:1707.00753 [hep-ph]}}.
\mciteBstWouldAddEndPunctfalse
\mciteSetBstMidEndSepPunct{\mcitedefaultmidpunct}
{}{\mcitedefaultseppunct}\relax
\EndOfBibitem
\bibitem{Dermisek:2013gta}
R.~Dermisek and A.~Raval, ``{Explanation of the Muon g-2 Anomaly with
  Vectorlike Leptons and its Implications for Higgs Decays},''
  \href{http://dx.doi.org/10.1103/PhysRevD.88.013017}{{\em Phys. Rev.}
  {\bfseries D88} (2013) 013017},
\href{http://arxiv.org/abs/1305.3522}{{\ttfamily arXiv:1305.3522 [hep-ph]}}.
\mciteBstWouldAddEndPunctfalse
\mciteSetBstMidEndSepPunct{\mcitedefaultmidpunct}
{}{\mcitedefaultseppunct}\relax
\EndOfBibitem
\bibitem{Freitas:2014pua}
A.~Freitas, J.~Lykken, S.~Kell, and S.~Westhoff, ``{Testing the Muon g-2
  Anomaly at the LHC},'' \href{http://dx.doi.org/10.1007/JHEP09(2014)155,
  10.1007/JHEP05(2014)145}{{\em JHEP} {\bfseries 05} (2014) 145},
  \href{http://arxiv.org/abs/1402.7065}{{\ttfamily arXiv:1402.7065 [hep-ph]}}.
[Erratum: JHEP09,155(2014)].
\mciteBstWouldAddEndPunctfalse
\mciteSetBstMidEndSepPunct{\mcitedefaultmidpunct}
{}{\mcitedefaultseppunct}\relax
\EndOfBibitem
\bibitem{ATLAS-CONF-2013-070}
``{Search for New Physics in Events with Three Charged Leptons with the ATLAS
  detector},'' Tech. Rep. ATLAS-CONF-2013-070, CERN, Geneva, Jul, 2013.
\newblock \url{http://cds.cern.ch/record/1562898}\relax
\mciteBstWouldAddEndPunctfalse
\mciteSetBstMidEndSepPunct{\mcitedefaultmidpunct}
{}{\mcitedefaultseppunct}\relax
\EndOfBibitem
\bibitem{Aad:2014hja}
{\bfseries ATLAS} Collaboration, G.~Aad {\em et~al.}, ``{Search for new
  phenomena in events with three or more charged leptons in $pp$ collisions at
  $\sqrt{s}=8$ TeV with the ATLAS detector},''
  \href{http://dx.doi.org/10.1007/JHEP08(2015)138}{{\em JHEP} {\bfseries 08}
  (2015) 138},
\href{http://arxiv.org/abs/1411.2921}{{\ttfamily arXiv:1411.2921 [hep-ex]}}.
\mciteBstWouldAddEndPunctfalse
\mciteSetBstMidEndSepPunct{\mcitedefaultmidpunct}
{}{\mcitedefaultseppunct}\relax
\EndOfBibitem
\bibitem{Djouadi:1989md}
A.~Djouadi, T.~Kohler, M.~Spira, and J.~Tutas, ``{(e b), (e t) TYPE LEPTOQUARKS
  AT e p COLLIDERS},''
\href{http://dx.doi.org/10.1007/BF01560270}{{\em Z. Phys.} {\bfseries C46}
  (1990) 679--686}.
\mciteBstWouldAddEndPunctfalse
\mciteSetBstMidEndSepPunct{\mcitedefaultmidpunct}
{}{\mcitedefaultseppunct}\relax
\EndOfBibitem
\bibitem{Chakraverty:2001yg}
D.~Chakraverty, D.~Choudhury, and A.~Datta, ``{A Nonsupersymmetric resolution
  of the anomalous muon magnetic moment},''
  \href{http://dx.doi.org/10.1016/S0370-2693(01)00419-1}{{\em Phys. Lett.}
  {\bfseries B506} (2001) 103--108},
\href{http://arxiv.org/abs/hep-ph/0102180}{{\ttfamily arXiv:hep-ph/0102180
  [hep-ph]}}.
\mciteBstWouldAddEndPunctfalse
\mciteSetBstMidEndSepPunct{\mcitedefaultmidpunct}
{}{\mcitedefaultseppunct}\relax
\EndOfBibitem
\bibitem{Cheung:2001ip}
K.-m. Cheung, ``{Muon anomalous magnetic moment and leptoquark solutions},''
  \href{http://dx.doi.org/10.1103/PhysRevD.64.033001}{{\em Phys. Rev.}
  {\bfseries D64} (2001) 033001},
\href{http://arxiv.org/abs/hep-ph/0102238}{{\ttfamily arXiv:hep-ph/0102238
  [hep-ph]}}.
\mciteBstWouldAddEndPunctfalse
\mciteSetBstMidEndSepPunct{\mcitedefaultmidpunct}
{}{\mcitedefaultseppunct}\relax
\EndOfBibitem
\bibitem{Queiroz:2014zfa}
F.~S. Queiroz and W.~Shepherd, ``{New Physics Contributions to the Muon
  Anomalous Magnetic Moment: A Numerical Code},''
  \href{http://dx.doi.org/10.1103/PhysRevD.89.095024}{{\em Phys. Rev.}
  {\bfseries D89} no.~9, (2014) 095024},
\href{http://arxiv.org/abs/1403.2309}{{\ttfamily arXiv:1403.2309 [hep-ph]}}.
\mciteBstWouldAddEndPunctfalse
\mciteSetBstMidEndSepPunct{\mcitedefaultmidpunct}
{}{\mcitedefaultseppunct}\relax
\EndOfBibitem
\bibitem{Sakaki:2013bfa}
Y.~Sakaki, M.~Tanaka, A.~Tayduganov, and R.~Watanabe, ``{Testing leptoquark
  models in $\bar B \to D^{(*)} \tau \bar\nu$},''
  \href{http://dx.doi.org/10.1103/PhysRevD.88.094012}{{\em Phys. Rev.}
  {\bfseries D88} no.~9, (2013) 094012},
\href{http://arxiv.org/abs/1309.0301}{{\ttfamily arXiv:1309.0301 [hep-ph]}}.
\mciteBstWouldAddEndPunctfalse
\mciteSetBstMidEndSepPunct{\mcitedefaultmidpunct}
{}{\mcitedefaultseppunct}\relax
\EndOfBibitem
\bibitem{Hiller:2014yaa}
G.~Hiller and M.~Schmaltz, ``{$R_K$ and future $b \to s \ell \ell$ physics
  beyond the standard model opportunities},''
  \href{http://dx.doi.org/10.1103/PhysRevD.90.054014}{{\em Phys. Rev.}
  {\bfseries D90} (2014) 054014},
\href{http://arxiv.org/abs/1408.1627}{{\ttfamily arXiv:1408.1627 [hep-ph]}}.
\mciteBstWouldAddEndPunctfalse
\mciteSetBstMidEndSepPunct{\mcitedefaultmidpunct}
{}{\mcitedefaultseppunct}\relax
\EndOfBibitem
\bibitem{Gripaios:2014tna}
B.~Gripaios, M.~Nardecchia, and S.~A. Renner, ``{Composite leptoquarks and
  anomalies in $B$-meson decays},''
  \href{http://dx.doi.org/10.1007/JHEP05(2015)006}{{\em JHEP} {\bfseries 05}
  (2015) 006},
\href{http://arxiv.org/abs/1412.1791}{{\ttfamily arXiv:1412.1791 [hep-ph]}}.
\mciteBstWouldAddEndPunctfalse
\mciteSetBstMidEndSepPunct{\mcitedefaultmidpunct}
{}{\mcitedefaultseppunct}\relax
\EndOfBibitem
\bibitem{Varzielas:2015iva}
I.~de~Medeiros~Varzielas and G.~Hiller, ``{Clues for flavor from rare lepton
  and quark decays},'' \href{http://dx.doi.org/10.1007/JHEP06(2015)072}{{\em
  JHEP} {\bfseries 06} (2015) 072},
\href{http://arxiv.org/abs/1503.01084}{{\ttfamily arXiv:1503.01084 [hep-ph]}}.
\mciteBstWouldAddEndPunctfalse
\mciteSetBstMidEndSepPunct{\mcitedefaultmidpunct}
{}{\mcitedefaultseppunct}\relax
\EndOfBibitem
\bibitem{Calibbi:2015kma}
L.~Calibbi, A.~Crivellin, and T.~Ota, ``{Effective Field Theory Approach to
  $b\rightarrow s l l'$, $B\rightarrow K^{(\ast)}\nu\bar{\nu}$ and
  $B\rightarrow D^{(\ast)}\tau\nu$ with Third Generation Couplings},''
  \href{http://dx.doi.org/10.1103/PhysRevLett.115.181801}{{\em Phys. Rev.
  Lett.} {\bfseries 115} (2015) 181801},
\href{http://arxiv.org/abs/1506.02661}{{\ttfamily arXiv:1506.02661 [hep-ph]}}.
\mciteBstWouldAddEndPunctfalse
\mciteSetBstMidEndSepPunct{\mcitedefaultmidpunct}
{}{\mcitedefaultseppunct}\relax
\EndOfBibitem
\bibitem{Freytsis:2015qca}
M.~Freytsis, Z.~Ligeti, and J.~T. Ruderman, ``{Flavor models for $\bar{B} \to
  D^{(*)} \tau \bar{\nu}$},''
  \href{http://dx.doi.org/10.1103/PhysRevD.92.054018}{{\em Phys. Rev.}
  {\bfseries D92} no.~5, (2015) 054018},
\href{http://arxiv.org/abs/1506.08896}{{\ttfamily arXiv:1506.08896 [hep-ph]}}.
\mciteBstWouldAddEndPunctfalse
\mciteSetBstMidEndSepPunct{\mcitedefaultmidpunct}
{}{\mcitedefaultseppunct}\relax
\EndOfBibitem
\bibitem{Bauer:2015knc}
M.~Bauer and M.~Neubert, ``{Minimal Leptoquark Explanation for the
  R$_{D^{(*)}}$ , R$_K$ , and $(g-2)_g$ Anomalies},''
  \href{http://dx.doi.org/10.1103/PhysRevLett.116.141802}{{\em Phys. Rev.
  Lett.} {\bfseries 116} no.~14, (2016) 141802},
\href{http://arxiv.org/abs/1511.01900}{{\ttfamily arXiv:1511.01900 [hep-ph]}}.
\mciteBstWouldAddEndPunctfalse
\mciteSetBstMidEndSepPunct{\mcitedefaultmidpunct}
{}{\mcitedefaultseppunct}\relax
\EndOfBibitem
\bibitem{Fajfer:2015ycq}
S.~Fajfer and N.~Ko{\v s}nik, ``{Vector leptoquark resolution of $R_K$ and
  $R_{D^{(*)}}$ puzzles},''
  \href{http://dx.doi.org/10.1016/j.physletb.2016.02.018}{{\em Phys. Lett.}
  {\bfseries B755} (2016) 270--274},
\href{http://arxiv.org/abs/1511.06024}{{\ttfamily arXiv:1511.06024 [hep-ph]}}.
\mciteBstWouldAddEndPunctfalse
\mciteSetBstMidEndSepPunct{\mcitedefaultmidpunct}
{}{\mcitedefaultseppunct}\relax
\EndOfBibitem
\bibitem{Barbieri:2015yvd}
R.~Barbieri, G.~Isidori, A.~Pattori, and F.~Senia, ``{Anomalies in $B$-decays
  and $U(2)$ flavour symmetry},''
  \href{http://dx.doi.org/10.1140/epjc/s10052-016-3905-3}{{\em Eur. Phys. J.}
  {\bfseries C76} no.~2, (2016) 67},
\href{http://arxiv.org/abs/1512.01560}{{\ttfamily arXiv:1512.01560 [hep-ph]}}.
\mciteBstWouldAddEndPunctfalse
\mciteSetBstMidEndSepPunct{\mcitedefaultmidpunct}
{}{\mcitedefaultseppunct}\relax
\EndOfBibitem
\bibitem{ColuccioLeskow:2016dox}
E.~Coluccio~Leskow, G.~D'Ambrosio, A.~Crivellin, and D.~M{\" u}ller,
  ``{$(g-2)\mu$, lepton flavor violation, and $Z$ decays with leptoquarks:
  Correlations and future prospects},''
  \href{http://dx.doi.org/10.1103/PhysRevD.95.055018}{{\em Phys. Rev.}
  {\bfseries D95} no.~5, (2017) 055018},
\href{http://arxiv.org/abs/1612.06858}{{\ttfamily arXiv:1612.06858 [hep-ph]}}.
\mciteBstWouldAddEndPunctfalse
\mciteSetBstMidEndSepPunct{\mcitedefaultmidpunct}
{}{\mcitedefaultseppunct}\relax
\EndOfBibitem
\bibitem{Bar-Shalom:2018ure}
S.~Bar-Shalom, J.~Cohen, A.~Soni, and J.~Wudka, ``{Phenomenology of TeV-scale
  scalar Leptoquarks in the EFT},''
\href{http://arxiv.org/abs/1812.03178}{{\ttfamily arXiv:1812.03178 [hep-ph]}}.
\mciteBstWouldAddEndPunctfalse
\mciteSetBstMidEndSepPunct{\mcitedefaultmidpunct}
{}{\mcitedefaultseppunct}\relax
\EndOfBibitem
\bibitem{Calibbi:2018rzv}
L.~Calibbi, R.~Ziegler, and J.~Zupan, ``{Minimal models for dark matter and the
  muon $g-2$ anomaly},'' \href{http://dx.doi.org/10.1007/JHEP07(2018)046}{{\em
  JHEP} {\bfseries 07} (2018) 046},
\href{http://arxiv.org/abs/1804.00009}{{\ttfamily arXiv:1804.00009 [hep-ph]}}.
\mciteBstWouldAddEndPunctfalse
\mciteSetBstMidEndSepPunct{\mcitedefaultmidpunct}
{}{\mcitedefaultseppunct}\relax
\EndOfBibitem
\bibitem{Aaboud:2017buh}
{\bfseries ATLAS} Collaboration, M.~Aaboud {\em et~al.}, ``{Search for new
  high-mass phenomena in the dilepton final state using 36 fb$^{−1}$ of
  proton-proton collision data at $ \sqrt{s}=13 $ TeV with the ATLAS
  detector},'' \href{http://dx.doi.org/10.1007/JHEP10(2017)182}{{\em JHEP}
  {\bfseries 10} (2017) 182},
\href{http://arxiv.org/abs/1707.02424}{{\ttfamily arXiv:1707.02424 [hep-ex]}}.
\mciteBstWouldAddEndPunctfalse
\mciteSetBstMidEndSepPunct{\mcitedefaultmidpunct}
{}{\mcitedefaultseppunct}\relax
\EndOfBibitem
\bibitem{Sirunyan:2018exx}
{\bfseries CMS} Collaboration, A.~M. Sirunyan {\em et~al.}, ``{Search for
  high-mass resonances in dilepton final states in proton-proton collisions at
  $\sqrt{s}=$ 13 TeV},'' \href{http://dx.doi.org/10.1007/JHEP06(2018)120}{{\em
  JHEP} {\bfseries 06} (2018) 120},
\href{http://arxiv.org/abs/1803.06292}{{\ttfamily arXiv:1803.06292 [hep-ex]}}.
\mciteBstWouldAddEndPunctfalse
\mciteSetBstMidEndSepPunct{\mcitedefaultmidpunct}
{}{\mcitedefaultseppunct}\relax
\EndOfBibitem
\bibitem{PhysRevLett.126.141801}
{\bfseries Muon $g-2$ Collaboration} Collaboration, B.~Abi {\em et~al.},
  ``Measurement of the positive muon anomalous magnetic moment to 0.46 ppm,''
  \href{http://dx.doi.org/10.1103/PhysRevLett.126.141801}{{\em Phys. Rev.
  Lett.} {\bfseries 126} (Apr, 2021) 141801}.
  \url{https://link.aps.org/doi/10.1103/PhysRevLett.126.141801}\relax
\mciteBstWouldAddEndPunctfalse
\mciteSetBstMidEndSepPunct{\mcitedefaultmidpunct}
{}{\mcitedefaultseppunct}\relax
\EndOfBibitem
\bibitem{Eichten:1986eq}
E.~Eichten, I.~Hinchliffe, K.~D. Lane, and C.~Quigg, ``{Signatures for
  Technicolor},''
\href{http://dx.doi.org/10.1103/PhysRevD.34.1547}{{\em Phys. Rev.} {\bfseries
  D34} (1986) 1547}.
\mciteBstWouldAddEndPunctfalse
\mciteSetBstMidEndSepPunct{\mcitedefaultmidpunct}
{}{\mcitedefaultseppunct}\relax
\EndOfBibitem
\bibitem{Lane:1991qh}
K.~D. Lane and M.~V. Ramana, ``{Walking technicolor signatures at hadron
  colliders},''
\href{http://dx.doi.org/10.1103/PhysRevD.44.2678}{{\em Phys. Rev.} {\bfseries
  D44} (1991) 2678--2700}.
\mciteBstWouldAddEndPunctfalse
\mciteSetBstMidEndSepPunct{\mcitedefaultmidpunct}
{}{\mcitedefaultseppunct}\relax
\EndOfBibitem
\bibitem{Assad:2017iib}
N.~Assad, B.~Fornal, and B.~Grinstein, ``{Baryon Number and Lepton Universality
  Violation in Leptoquark and Diquark Models},''
  \href{http://dx.doi.org/10.1016/j.physletb.2017.12.042}{{\em Phys. Lett.}
  {\bfseries B777} (2018) 324--331},
\href{http://arxiv.org/abs/1708.06350}{{\ttfamily arXiv:1708.06350 [hep-ph]}}.
\mciteBstWouldAddEndPunctfalse
\mciteSetBstMidEndSepPunct{\mcitedefaultmidpunct}
{}{\mcitedefaultseppunct}\relax
\EndOfBibitem
\bibitem{DiLuzio:2017vat}
L.~Di~Luzio, A.~Greljo, and M.~Nardecchia, ``{Gauge leptoquark as the origin of
  B-physics anomalies},''
  \href{http://dx.doi.org/10.1103/PhysRevD.96.115011}{{\em Phys. Rev.}
  {\bfseries D96} no.~11, (2017) 115011},
\href{http://arxiv.org/abs/1708.08450}{{\ttfamily arXiv:1708.08450 [hep-ph]}}.
\mciteBstWouldAddEndPunctfalse
\mciteSetBstMidEndSepPunct{\mcitedefaultmidpunct}
{}{\mcitedefaultseppunct}\relax
\EndOfBibitem
\bibitem{Calibbi:2017qbu}
L.~Calibbi, A.~Crivellin, and T.~Li, ``{A model of vector leptoquarks in view
  of the $B$-physics anomalies},''
  \href{http://dx.doi.org/10.1103/PhysRevD.98.115002}{{\em Phys. Rev.}
  {\bfseries D98} (2018) 115002},
\href{http://arxiv.org/abs/1709.00692}{{\ttfamily arXiv:1709.00692 [hep-ph]}}.
\mciteBstWouldAddEndPunctfalse
\mciteSetBstMidEndSepPunct{\mcitedefaultmidpunct}
{}{\mcitedefaultseppunct}\relax
\EndOfBibitem
\bibitem{Bordone:2017bld}
M.~Bordone, C.~Cornella, J.~Fuentes-Martin, and G.~Isidori, ``{A three-site
  gauge model for flavor hierarchies and flavor anomalies},''
  \href{http://dx.doi.org/10.1016/j.physletb.2018.02.011}{{\em Phys. Lett.}
  {\bfseries B779} (2018) 317--323},
\href{http://arxiv.org/abs/1712.01368}{{\ttfamily arXiv:1712.01368 [hep-ph]}}.
\mciteBstWouldAddEndPunctfalse
\mciteSetBstMidEndSepPunct{\mcitedefaultmidpunct}
{}{\mcitedefaultseppunct}\relax
\EndOfBibitem
\bibitem{Fornal:2018dqn}
B.~Fornal, S.~A. Gadam, and B.~Grinstein, ``{Left-Right SU(4) Vector Leptoquark
  Model for Flavor Anomalies},''
\href{http://arxiv.org/abs/1812.01603}{{\ttfamily arXiv:1812.01603 [hep-ph]}}.
\mciteBstWouldAddEndPunctfalse
\mciteSetBstMidEndSepPunct{\mcitedefaultmidpunct}
{}{\mcitedefaultseppunct}\relax
\EndOfBibitem
\bibitem{Sirunyan:2018kzh}
{\bfseries CMS} Collaboration, A.~M. Sirunyan {\em et~al.}, ``{Constraints on
  models of scalar and vector leptoquarks decaying to a quark and a neutrino at
  $\sqrt{s}=$ 13 TeV},''
  \href{http://dx.doi.org/10.1103/PhysRevD.98.032005}{{\em Phys. Rev.}
  {\bfseries D98} no.~3, (2018) 032005},
\href{http://arxiv.org/abs/1805.10228}{{\ttfamily arXiv:1805.10228 [hep-ex]}}.
\mciteBstWouldAddEndPunctfalse
\mciteSetBstMidEndSepPunct{\mcitedefaultmidpunct}
{}{\mcitedefaultseppunct}\relax
\EndOfBibitem
\bibitem{Schmaltz:2018nls}
M.~Schmaltz and Y.-M. Zhong, ``{The leptoquark Hunter's guide: large
  coupling},'' \href{http://dx.doi.org/10.1007/JHEP01(2019)132}{{\em JHEP}
  {\bfseries 01} (2019) 132},
\href{http://arxiv.org/abs/1810.10017}{{\ttfamily arXiv:1810.10017 [hep-ph]}}.
\mciteBstWouldAddEndPunctfalse
\mciteSetBstMidEndSepPunct{\mcitedefaultmidpunct}
{}{\mcitedefaultseppunct}\relax
\EndOfBibitem
\bibitem{Angelescu:2018tyl}
A.~Angelescu, D.~Bečirević, D.~A. Faroughy, and O.~Sumensari, ``{Closing the
  window on single leptoquark solutions to the $B$-physics anomalies},''
  \href{http://dx.doi.org/10.1007/JHEP10(2018)183}{{\em JHEP} {\bfseries 10}
  (2018) 183},
\href{http://arxiv.org/abs/1808.08179}{{\ttfamily arXiv:1808.08179 [hep-ph]}}.
\mciteBstWouldAddEndPunctfalse
\mciteSetBstMidEndSepPunct{\mcitedefaultmidpunct}
{}{\mcitedefaultseppunct}\relax
\EndOfBibitem
\bibitem{Keith:1997fv}
E.~Keith and E.~Ma, ``{S, T, and leptoquarks at HERA},''
  \href{http://dx.doi.org/10.1103/PhysRevLett.79.4318}{{\em Phys. Rev. Lett.}
  {\bfseries 79} (1997) 4318--4320},
\href{http://arxiv.org/abs/hep-ph/9707214}{{\ttfamily arXiv:hep-ph/9707214
  [hep-ph]}}.
\mciteBstWouldAddEndPunctfalse
\mciteSetBstMidEndSepPunct{\mcitedefaultmidpunct}
{}{\mcitedefaultseppunct}\relax
\EndOfBibitem
\bibitem{Baak:2014ora}
{\bfseries Gfitter Group} Collaboration, M.~Baak, J.~Cúth, J.~Haller,
  A.~Hoecker, R.~Kogler, K.~Mönig, M.~Schott, and J.~Stelzer, ``{The global
  electroweak fit at NNLO and prospects for the LHC and ILC},''
  \href{http://dx.doi.org/10.1140/epjc/s10052-014-3046-5}{{\em Eur. Phys. J.}
  {\bfseries C74} (2014) 3046},
\href{http://arxiv.org/abs/1407.3792}{{\ttfamily arXiv:1407.3792 [hep-ph]}}.
\mciteBstWouldAddEndPunctfalse
\mciteSetBstMidEndSepPunct{\mcitedefaultmidpunct}
{}{\mcitedefaultseppunct}\relax
\EndOfBibitem
\bibitem{deBoer:2015boa}
S.~de~Boer and G.~Hiller, ``{Flavor and new physics opportunities with rare
  charm decays into leptons},''
  \href{http://dx.doi.org/10.1103/PhysRevD.93.074001}{{\em Phys. Rev.}
  {\bfseries D93} no.~7, (2016) 074001},
\href{http://arxiv.org/abs/1510.00311}{{\ttfamily arXiv:1510.00311 [hep-ph]}}.
\mciteBstWouldAddEndPunctfalse
\mciteSetBstMidEndSepPunct{\mcitedefaultmidpunct}
{}{\mcitedefaultseppunct}\relax
\EndOfBibitem
\bibitem{Aaij:2013cza}
{\bfseries LHCb} Collaboration, R.~Aaij {\em et~al.}, ``{Search for the rare
  decay $D^0 \to \mu^+ \mu^-$},''
  \href{http://dx.doi.org/10.1016/j.physletb.2013.06.037}{{\em Phys. Lett.}
  {\bfseries B725} (2013) 15--24},
\href{http://arxiv.org/abs/1305.5059}{{\ttfamily arXiv:1305.5059 [hep-ex]}}.
\mciteBstWouldAddEndPunctfalse
\mciteSetBstMidEndSepPunct{\mcitedefaultmidpunct}
{}{\mcitedefaultseppunct}\relax
\EndOfBibitem
\bibitem{Fajfer:2015mia}
S.~Fajfer and N.~Ko{\v s}nik, ``{Prospects of discovering new physics in rare
  charm decays},'' \href{http://dx.doi.org/10.1140/epjc/s10052-015-3801-2}{{\em
  Eur. Phys. J.} {\bfseries C75} no.~12, (2015) 567},
\href{http://arxiv.org/abs/1510.00965}{{\ttfamily arXiv:1510.00965 [hep-ph]}}.
\mciteBstWouldAddEndPunctfalse
\mciteSetBstMidEndSepPunct{\mcitedefaultmidpunct}
{}{\mcitedefaultseppunct}\relax
\EndOfBibitem
\bibitem{Cai:2017wry}
Y.~Cai, J.~Gargalionis, M.~A. Schmidt, and R.~R. Volkas, ``{Reconsidering the
  One Leptoquark solution: flavor anomalies and neutrino mass},''
  \href{http://dx.doi.org/10.1007/JHEP10(2017)047}{{\em JHEP} {\bfseries 10}
  (2017) 047},
\href{http://arxiv.org/abs/1704.05849}{{\ttfamily arXiv:1704.05849 [hep-ph]}}.
\mciteBstWouldAddEndPunctfalse
\mciteSetBstMidEndSepPunct{\mcitedefaultmidpunct}
{}{\mcitedefaultseppunct}\relax
\EndOfBibitem
\bibitem{Paul:2010pq}
A.~Paul, I.~I. Bigi, and S.~Recksiegel, ``{$D^0 \to \gamma \gamma$ and $D^0 \to
  \mu^+ \mu^-$ Rates on an Unlikely Impact of the Littlest Higgs Model with
  T-Parity},'' \href{http://dx.doi.org/10.1103/PhysRevD.83.019901,
  10.1103/PhysRevD.82.094006}{{\em Phys. Rev.} {\bfseries D82} (2010) 094006},
  \href{http://arxiv.org/abs/1008.3141}{{\ttfamily arXiv:1008.3141 [hep-ph]}}.
[Erratum: Phys. Rev.D83,019901(2011)].
\mciteBstWouldAddEndPunctfalse
\mciteSetBstMidEndSepPunct{\mcitedefaultmidpunct}
{}{\mcitedefaultseppunct}\relax
\EndOfBibitem
\bibitem{Mohr:2015ccw}
P.~J. Mohr, D.~B. Newell, and B.~N. Taylor, ``{CODATA Recommended Values of the
  Fundamental Physical Constants: 2014},''
  \href{http://dx.doi.org/10.1103/RevModPhys.88.035009}{{\em Rev. Mod. Phys.}
  {\bfseries 88} no.~3, (2016) 035009},
\href{http://arxiv.org/abs/1507.07956}{{\ttfamily arXiv:1507.07956
  [physics.atom-ph]}}.
\mciteBstWouldAddEndPunctfalse
\mciteSetBstMidEndSepPunct{\mcitedefaultmidpunct}
{}{\mcitedefaultseppunct}\relax
\EndOfBibitem
\bibitem{Tanabashi:2018oca}
{\bfseries Particle Data Group} Collaboration, M.~Tanabashi {\em et~al.},
  ``{Review of Particle Physics},''
\href{http://dx.doi.org/10.1103/PhysRevD.98.030001}{{\em Phys. Rev.} {\bfseries
  D98} no.~3, (2018) 030001}.
\mciteBstWouldAddEndPunctfalse
\mciteSetBstMidEndSepPunct{\mcitedefaultmidpunct}
{}{\mcitedefaultseppunct}\relax
\EndOfBibitem
\bibitem{Lees:2013uzd}
{\bfseries BaBar} Collaboration, J.~P. Lees {\em et~al.}, ``{Measurement of an
  Excess of $\bar{B} \to D^{(*)}\tau^- \bar{\nu}_\tau$ Decays and Implications
  for Charged Higgs Bosons},''
  \href{http://dx.doi.org/10.1103/PhysRevD.88.072012}{{\em Phys. Rev.}
  {\bfseries D88} no.~7, (2013) 072012},
\href{http://arxiv.org/abs/1303.0571}{{\ttfamily arXiv:1303.0571 [hep-ex]}}.
\mciteBstWouldAddEndPunctfalse
\mciteSetBstMidEndSepPunct{\mcitedefaultmidpunct}
{}{\mcitedefaultseppunct}\relax
\EndOfBibitem
\bibitem{Aaij:2015yra}
{\bfseries LHCb} Collaboration, R.~Aaij {\em et~al.}, ``{Measurement of the
  ratio of branching fractions $\mathcal{B}(\bar{B}^0 \to
  D^{*+}\tau^{-}\bar{\nu}_{\tau})/\mathcal{B}(\bar{B}^0 \to
  D^{*+}\mu^{-}\bar{\nu}_{\mu})$},''
  \href{http://dx.doi.org/10.1103/PhysRevLett.115.159901,
  10.1103/PhysRevLett.115.111803}{{\em Phys. Rev. Lett.} {\bfseries 115}
  no.~11, (2015) 111803}, \href{http://arxiv.org/abs/1506.08614}{{\ttfamily
  arXiv:1506.08614 [hep-ex]}}.
[Erratum: Phys. Rev. Lett.115,no.15,159901(2015)].
\mciteBstWouldAddEndPunctfalse
\mciteSetBstMidEndSepPunct{\mcitedefaultmidpunct}
{}{\mcitedefaultseppunct}\relax
\EndOfBibitem
\bibitem{Hirose:2016wfn}
{\bfseries Belle} Collaboration, S.~Hirose {\em et~al.}, ``{Measurement of the
  $\tau$ lepton polarization and $R(D^*)$ in the decay $\bar{B} \to D^* \tau^-
  \bar{\nu}_\tau$},''
  \href{http://dx.doi.org/10.1103/PhysRevLett.118.211801}{{\em Phys. Rev.
  Lett.} {\bfseries 118} no.~21, (2017) 211801},
\href{http://arxiv.org/abs/1612.00529}{{\ttfamily arXiv:1612.00529 [hep-ex]}}.
\mciteBstWouldAddEndPunctfalse
\mciteSetBstMidEndSepPunct{\mcitedefaultmidpunct}
{}{\mcitedefaultseppunct}\relax
\EndOfBibitem
\bibitem{Aaij:2014ora}
{\bfseries LHCb} Collaboration, R.~Aaij {\em et~al.}, ``{Test of lepton
  universality using $B^{+}\rightarrow K^{+}\ell^{+}\ell^{-}$ decays},''
  \href{http://dx.doi.org/10.1103/PhysRevLett.113.151601}{{\em Phys. Rev.
  Lett.} {\bfseries 113} (2014) 151601},
\href{http://arxiv.org/abs/1406.6482}{{\ttfamily arXiv:1406.6482 [hep-ex]}}.
\mciteBstWouldAddEndPunctfalse
\mciteSetBstMidEndSepPunct{\mcitedefaultmidpunct}
{}{\mcitedefaultseppunct}\relax
\EndOfBibitem
\bibitem{Aaij:2017vbb}
{\bfseries LHCb} Collaboration, R.~Aaij {\em et~al.}, ``{Test of lepton
  universality with $B^{0} \rightarrow K^{*0}\ell^{+}\ell^{-}$ decays},''
  \href{http://dx.doi.org/10.1007/JHEP08(2017)055}{{\em JHEP} {\bfseries 08}
  (2017) 055},
\href{http://arxiv.org/abs/1705.05802}{{\ttfamily arXiv:1705.05802 [hep-ex]}}.
\mciteBstWouldAddEndPunctfalse
\mciteSetBstMidEndSepPunct{\mcitedefaultmidpunct}
{}{\mcitedefaultseppunct}\relax
\EndOfBibitem
\bibitem{Durieux:2014xla}
G.~Durieux, F.~Maltoni, and C.~Zhang, ``{Global approach to top-quark
  flavor-changing interactions},''
  \href{http://dx.doi.org/10.1103/PhysRevD.91.074017}{{\em Phys. Rev.}
  {\bfseries D91} no.~7, (2015) 074017},
\href{http://arxiv.org/abs/1412.7166}{{\ttfamily arXiv:1412.7166 [hep-ph]}}.
\mciteBstWouldAddEndPunctfalse
\mciteSetBstMidEndSepPunct{\mcitedefaultmidpunct}
{}{\mcitedefaultseppunct}\relax
\EndOfBibitem
\bibitem{Chala:2018agk}
M.~Chala, J.~Santiago, and M.~Spannowsky, ``{Constraining four-fermion
  operators using rare top decays},''
\href{http://arxiv.org/abs/1809.09624}{{\ttfamily arXiv:1809.09624 [hep-ph]}}.
\mciteBstWouldAddEndPunctfalse
\mciteSetBstMidEndSepPunct{\mcitedefaultmidpunct}
{}{\mcitedefaultseppunct}\relax
\EndOfBibitem
\bibitem{Becirevic:2017jtw}
D.~Be{\v c}irevi{\'c} and O.~Sumensari, ``{A leptoquark model to accommodate
  $R_K^\mathrm{exp} < R_K^\mathrm{SM}$ and $R_{K^\ast}^\mathrm{exp} <
  R_{K^\ast}^\mathrm{SM}$},''
  \href{http://dx.doi.org/10.1007/JHEP08(2017)104}{{\em JHEP} {\bfseries 08}
  (2017) 104},
\href{http://arxiv.org/abs/1704.05835}{{\ttfamily arXiv:1704.05835 [hep-ph]}}.
\mciteBstWouldAddEndPunctfalse
\mciteSetBstMidEndSepPunct{\mcitedefaultmidpunct}
{}{\mcitedefaultseppunct}\relax
\EndOfBibitem
\bibitem{Buras:2014fpa}
A.~J. Buras, J.~Girrbach-Noe, C.~Niehoff, and D.~M. Straub, ``{$ B\to
  {K}^{\left(\ast \right)}\nu \overline{\nu} $ decays in the Standard Model and
  beyond},'' \href{http://dx.doi.org/10.1007/JHEP02(2015)184}{{\em JHEP}
  {\bfseries 02} (2015) 184},
\href{http://arxiv.org/abs/1409.4557}{{\ttfamily arXiv:1409.4557 [hep-ph]}}.
\mciteBstWouldAddEndPunctfalse
\mciteSetBstMidEndSepPunct{\mcitedefaultmidpunct}
{}{\mcitedefaultseppunct}\relax
\EndOfBibitem
\bibitem{Kumar:2016omp}
G.~Kumar, ``{Constraints on a scalar leptoquark from the kaon sector},''
  \href{http://dx.doi.org/10.1103/PhysRevD.94.014022}{{\em Phys. Rev.}
  {\bfseries D94} no.~1, (2016) 014022},
\href{http://arxiv.org/abs/1603.00346}{{\ttfamily arXiv:1603.00346 [hep-ph]}}.
\mciteBstWouldAddEndPunctfalse
\mciteSetBstMidEndSepPunct{\mcitedefaultmidpunct}
{}{\mcitedefaultseppunct}\relax
\EndOfBibitem
\bibitem{Artamonov:2008qb}
{\bfseries E949} Collaboration, A.~V. Artamonov {\em et~al.}, ``{New
  measurement of the $K^{+} \to \pi^{+} \nu \bar{\nu}$ branching ratio},''
  \href{http://dx.doi.org/10.1103/PhysRevLett.101.191802}{{\em Phys. Rev.
  Lett.} {\bfseries 101} (2008) 191802},
\href{http://arxiv.org/abs/0808.2459}{{\ttfamily arXiv:0808.2459 [hep-ex]}}.
\mciteBstWouldAddEndPunctfalse
\mciteSetBstMidEndSepPunct{\mcitedefaultmidpunct}
{}{\mcitedefaultseppunct}\relax
\EndOfBibitem
\bibitem{Buchalla:1995vs}
G.~Buchalla, A.~J. Buras, and M.~E. Lautenbacher, ``{Weak decays beyond leading
  logarithms},'' \href{http://dx.doi.org/10.1103/RevModPhys.68.1125}{{\em Rev.
  Mod. Phys.} {\bfseries 68} (1996) 1125--1144},
\href{http://arxiv.org/abs/hep-ph/9512380}{{\ttfamily arXiv:hep-ph/9512380
  [hep-ph]}}.
\mciteBstWouldAddEndPunctfalse
\mciteSetBstMidEndSepPunct{\mcitedefaultmidpunct}
{}{\mcitedefaultseppunct}\relax
\EndOfBibitem
\bibitem{Brod:2008ss}
J.~Brod and M.~Gorbahn, ``{Electroweak Corrections to the Charm Quark
  Contribution to $K^+\to \pi^+ \nu \bar{\nu}$},''
  \href{http://dx.doi.org/10.1103/PhysRevD.78.034006}{{\em Phys. Rev.}
  {\bfseries D78} (2008) 034006},
\href{http://arxiv.org/abs/0805.4119}{{\ttfamily arXiv:0805.4119 [hep-ph]}}.
\mciteBstWouldAddEndPunctfalse
\mciteSetBstMidEndSepPunct{\mcitedefaultmidpunct}
{}{\mcitedefaultseppunct}\relax
\EndOfBibitem
\bibitem{0902.0160}
W.~Altmannshofer, A.~J. Buras, D.~M. Straub, and M.~Wick, ``{New strategies for
  New Physics search in $B \to K^{*} \nu \bar{\nu}$, $B \to K \nu \bar{\nu}$
  and $B \to X_{s} \nu \bar{\nu}$ decays},''
  \href{http://dx.doi.org/10.1088/1126-6708/2009/04/022}{{\em JHEP} {\bfseries
  04} (2009) 022},
\href{http://arxiv.org/abs/0902.0160}{{\ttfamily arXiv:0902.0160 [hep-ph]}}.
\mciteBstWouldAddEndPunctfalse
\mciteSetBstMidEndSepPunct{\mcitedefaultmidpunct}
{}{\mcitedefaultseppunct}\relax
\EndOfBibitem
\bibitem{Isidori:2003ts}
G.~Isidori and R.~Unterdorfer, ``{On the short distance constraints from
  $K_{L,S}\to \mu^+\mu^-$},''
  \href{http://dx.doi.org/10.1088/1126-6708/2004/01/009}{{\em JHEP} {\bfseries
  01} (2004) 009},
\href{http://arxiv.org/abs/hep-ph/0311084}{{\ttfamily arXiv:hep-ph/0311084
  [hep-ph]}}.
\mciteBstWouldAddEndPunctfalse
\mciteSetBstMidEndSepPunct{\mcitedefaultmidpunct}
{}{\mcitedefaultseppunct}\relax
\EndOfBibitem
\bibitem{Bansal:2018nwp}
S.~Bansal, R.~M. Capdevilla, and C.~Kolda, ``{Constraining the minimal flavor
  violating leptoquark explanation of the $R_{D^{(*)}}$ anomaly},''
  \href{http://dx.doi.org/10.1103/PhysRevD.99.035047}{{\em Phys. Rev.}
  {\bfseries D99} no.~3, (2019) 035047},
\href{http://arxiv.org/abs/1810.11588}{{\ttfamily arXiv:1810.11588 [hep-ph]}}.
\mciteBstWouldAddEndPunctfalse
\mciteSetBstMidEndSepPunct{\mcitedefaultmidpunct}
{}{\mcitedefaultseppunct}\relax
\EndOfBibitem
\bibitem{Aaboud:2016qeg}
{\bfseries ATLAS} Collaboration, M.~Aaboud {\em et~al.}, ``{Search for scalar
  leptoquarks in pp collisions at $\sqrt{s}$ = 13 TeV with the ATLAS
  experiment},'' \href{http://dx.doi.org/10.1088/1367-2630/18/9/093016}{{\em
  New J. Phys.} {\bfseries 18} no.~9, (2016) 093016},
\href{http://arxiv.org/abs/1605.06035}{{\ttfamily arXiv:1605.06035 [hep-ex]}}.
\mciteBstWouldAddEndPunctfalse
\mciteSetBstMidEndSepPunct{\mcitedefaultmidpunct}
{}{\mcitedefaultseppunct}\relax
\EndOfBibitem
\bibitem{Diaz:2017lit}
B.~Diaz, M.~Schmaltz, and Y.-M. Zhong, ``{The leptoquark Hunter's guide: Pair
  production},'' \href{http://dx.doi.org/10.1007/JHEP10(2017)097}{{\em JHEP}
  {\bfseries 10} (2017) 097},
\href{http://arxiv.org/abs/1706.05033}{{\ttfamily arXiv:1706.05033 [hep-ph]}}.
\mciteBstWouldAddEndPunctfalse
\mciteSetBstMidEndSepPunct{\mcitedefaultmidpunct}
{}{\mcitedefaultseppunct}\relax
\EndOfBibitem
\bibitem{Sirunyan:2018ryt}
{\bfseries CMS} Collaboration, A.~M. Sirunyan {\em et~al.}, ``{Search for pair
  production of second-generation leptoquarks at $\sqrt{s}=$ 13 TeV},''
  \href{http://dx.doi.org/10.1103/PhysRevD.99.032014}{{\em Phys. Rev.}
  {\bfseries D99} no.~3, (2019) 032014},
\href{http://arxiv.org/abs/1808.05082}{{\ttfamily arXiv:1808.05082 [hep-ex]}}.
\mciteBstWouldAddEndPunctfalse
\mciteSetBstMidEndSepPunct{\mcitedefaultmidpunct}
{}{\mcitedefaultseppunct}\relax
\EndOfBibitem
\bibitem{Faroughy:2016osc}
D.~A. Faroughy, A.~Greljo, and J.~F. Kamenik, ``{Confronting lepton flavor
  universality violation in B decays with high-$p_T$ tau lepton searches at
  LHC},'' \href{http://dx.doi.org/10.1016/j.physletb.2016.11.011}{{\em Phys.
  Lett.} {\bfseries B764} (2017) 126--134},
\href{http://arxiv.org/abs/1609.07138}{{\ttfamily arXiv:1609.07138 [hep-ph]}}.
\mciteBstWouldAddEndPunctfalse
\mciteSetBstMidEndSepPunct{\mcitedefaultmidpunct}
{}{\mcitedefaultseppunct}\relax
\EndOfBibitem
\bibitem{Greljo:2017vvb}
A.~Greljo and D.~Marzocca, ``{High-$p_T$ dilepton tails and flavor physics},''
  \href{http://dx.doi.org/10.1140/epjc/s10052-017-5119-8}{{\em Eur. Phys. J.}
  {\bfseries C77} no.~8, (2017) 548},
\href{http://arxiv.org/abs/1704.09015}{{\ttfamily arXiv:1704.09015 [hep-ph]}}.
\mciteBstWouldAddEndPunctfalse
\mciteSetBstMidEndSepPunct{\mcitedefaultmidpunct}
{}{\mcitedefaultseppunct}\relax
\EndOfBibitem
\bibitem{Bansal:2018eha}
S.~Bansal, R.~M. Capdevilla, A.~Delgado, C.~Kolda, A.~Martin, and N.~Raj,
  ``{Hunting leptoquarks in monolepton searches},''
  \href{http://dx.doi.org/10.1103/PhysRevD.98.015037}{{\em Phys. Rev.}
  {\bfseries D98} no.~1, (2018) 015037},
\href{http://arxiv.org/abs/1806.02370}{{\ttfamily arXiv:1806.02370 [hep-ph]}}.
\mciteBstWouldAddEndPunctfalse
\mciteSetBstMidEndSepPunct{\mcitedefaultmidpunct}
{}{\mcitedefaultseppunct}\relax
\EndOfBibitem
\bibitem{Raj:2016aky}
N.~Raj, ``{Anticipating nonresonant new physics in dilepton angular spectra at
  the LHC},'' \href{http://dx.doi.org/10.1103/PhysRevD.95.015011}{{\em Phys.
  Rev.} {\bfseries D95} no.~1, (2017) 015011},
\href{http://arxiv.org/abs/1610.03795}{{\ttfamily arXiv:1610.03795 [hep-ph]}}.
\mciteBstWouldAddEndPunctfalse
\mciteSetBstMidEndSepPunct{\mcitedefaultmidpunct}
{}{\mcitedefaultseppunct}\relax
\EndOfBibitem
\bibitem{Alloul:2013bka}
A.~Alloul, N.~D. Christensen, C.~Degrande, C.~Duhr, and B.~Fuks, ``{FeynRules
  2.0 - A complete toolbox for tree-level phenomenology},''
  \href{http://dx.doi.org/10.1016/j.cpc.2014.04.012}{{\em Comput. Phys.
  Commun.} {\bfseries 185} (2014) 2250--2300},
\href{http://arxiv.org/abs/1310.1921}{{\ttfamily arXiv:1310.1921 [hep-ph]}}.
\mciteBstWouldAddEndPunctfalse
\mciteSetBstMidEndSepPunct{\mcitedefaultmidpunct}
{}{\mcitedefaultseppunct}\relax
\EndOfBibitem
\bibitem{Alwall:2014hca}
J.~Alwall, R.~Frederix, S.~Frixione, V.~Hirschi, F.~Maltoni, O.~Mattelaer,
  H.~S. Shao, T.~Stelzer, P.~Torrielli, and M.~Zaro, ``{The automated
  computation of tree-level and next-to-leading order differential cross
  sections, and their matching to parton shower simulations},''
  \href{http://dx.doi.org/10.1007/JHEP07(2014)079}{{\em JHEP} {\bfseries 07}
  (2014) 079},
\href{http://arxiv.org/abs/1405.0301}{{\ttfamily arXiv:1405.0301 [hep-ph]}}.
\mciteBstWouldAddEndPunctfalse
\mciteSetBstMidEndSepPunct{\mcitedefaultmidpunct}
{}{\mcitedefaultseppunct}\relax
\EndOfBibitem
\bibitem{Porod:2003um}
W.~Porod, ``{SPheno, a program for calculating supersymmetric spectra, SUSY
  particle decays and SUSY particle production at e+ e- colliders},''
  \href{http://dx.doi.org/10.1016/S0010-4655(03)00222-4}{{\em Comput. Phys.
  Commun.} {\bfseries 153} (2003) 275--315},
\href{http://arxiv.org/abs/hep-ph/0301101}{{\ttfamily arXiv:hep-ph/0301101
  [hep-ph]}}.
\mciteBstWouldAddEndPunctfalse
\mciteSetBstMidEndSepPunct{\mcitedefaultmidpunct}
{}{\mcitedefaultseppunct}\relax
\EndOfBibitem
\bibitem{Porod:2011nf}
W.~Porod and F.~Staub, ``{SPheno 3.1: Extensions including flavour, CP-phases
  and models beyond the MSSM},''
  \href{http://dx.doi.org/10.1016/j.cpc.2012.05.021}{{\em Comput. Phys.
  Commun.} {\bfseries 183} (2012) 2458--2469},
\href{http://arxiv.org/abs/1104.1573}{{\ttfamily arXiv:1104.1573 [hep-ph]}}.
\mciteBstWouldAddEndPunctfalse
\mciteSetBstMidEndSepPunct{\mcitedefaultmidpunct}
{}{\mcitedefaultseppunct}\relax
\EndOfBibitem
\bibitem{Sjostrand:2007gs}
T.~Sjostrand, S.~Mrenna, and P.~Z. Skands, ``{A Brief Introduction to PYTHIA
  8.1},'' \href{http://dx.doi.org/10.1016/j.cpc.2008.01.036}{{\em Comput. Phys.
  Commun.} {\bfseries 178} (2008) 852--867},
\href{http://arxiv.org/abs/0710.3820}{{\ttfamily arXiv:0710.3820 [hep-ph]}}.
\mciteBstWouldAddEndPunctfalse
\mciteSetBstMidEndSepPunct{\mcitedefaultmidpunct}
{}{\mcitedefaultseppunct}\relax
\EndOfBibitem
\bibitem{deFavereau:2013fsa}
{\bfseries DELPHES 3} Collaboration, J.~de~Favereau, C.~Delaere, P.~Demin,
  A.~Giammanco, V.~Lemaître, A.~Mertens, and M.~Selvaggi, ``{DELPHES 3, A
  modular framework for fast simulation of a generic collider experiment},''
  \href{http://dx.doi.org/10.1007/JHEP02(2014)057}{{\em JHEP} {\bfseries 02}
  (2014) 057},
\href{http://arxiv.org/abs/1307.6346}{{\ttfamily arXiv:1307.6346 [hep-ex]}}.
\mciteBstWouldAddEndPunctfalse
\mciteSetBstMidEndSepPunct{\mcitedefaultmidpunct}
{}{\mcitedefaultseppunct}\relax
\EndOfBibitem
\bibitem{Anelli:2005ju}
G.~Anelli {\em et~al.},
``{Proposal to measure the rare decay $K^+\to\pi^+\nu\bar{\nu}$ at the CERN
  SPS},''.
\mciteBstWouldAddEndPunctfalse
\mciteSetBstMidEndSepPunct{\mcitedefaultmidpunct}
{}{\mcitedefaultseppunct}\relax
\EndOfBibitem
\bibitem{Davier:2017zfy}
M.~Davier, A.~Hoecker, B.~Malaescu, and Z.~Zhang, ``{Reevaluation of the
  hadronic vacuum polarisation contributions to the Standard Model predictions
  of the muon $g-2$ and ${\alpha (m_Z^2)}$ using newest hadronic cross-section
  data},'' \href{http://dx.doi.org/10.1140/epjc/s10052-017-5161-6}{{\em Eur.
  Phys. J. C} {\bfseries 77} no.~12, (2017) 827},
  \href{http://arxiv.org/abs/1706.09436}{{\ttfamily arXiv:1706.09436
  [hep-ph]}}\relax
\mciteBstWouldAddEndPunctfalse
\mciteSetBstMidEndSepPunct{\mcitedefaultmidpunct}
{}{\mcitedefaultseppunct}\relax
\EndOfBibitem
\bibitem{Keshavarzi:2018mgv}
A.~Keshavarzi, D.~Nomura, and T.~Teubner, ``{Muon $g-2$ and $\alpha(M_Z^2)$: a
  new data-based analysis},''
  \href{http://dx.doi.org/10.1103/PhysRevD.97.114025}{{\em Phys. Rev. D}
  {\bfseries 97} no.~11, (2018) 114025},
  \href{http://arxiv.org/abs/1802.02995}{{\ttfamily arXiv:1802.02995
  [hep-ph]}}\relax
\mciteBstWouldAddEndPunctfalse
\mciteSetBstMidEndSepPunct{\mcitedefaultmidpunct}
{}{\mcitedefaultseppunct}\relax
\EndOfBibitem
\bibitem{Colangelo:2018mtw}
G.~Colangelo, M.~Hoferichter, and P.~Stoffer, ``{Two-pion contribution to
  hadronic vacuum polarization},''
  \href{http://dx.doi.org/10.1007/JHEP02(2019)006}{{\em JHEP} {\bfseries 02}
  (2019) 006}, \href{http://arxiv.org/abs/1810.00007}{{\ttfamily
  arXiv:1810.00007 [hep-ph]}}\relax
\mciteBstWouldAddEndPunctfalse
\mciteSetBstMidEndSepPunct{\mcitedefaultmidpunct}
{}{\mcitedefaultseppunct}\relax
\EndOfBibitem
\bibitem{Hoferichter:2019mqg}
M.~Hoferichter, B.-L. Hoid, and B.~Kubis, ``{Three-pion contribution to
  hadronic vacuum polarization},''
  \href{http://dx.doi.org/10.1007/JHEP08(2019)137}{{\em JHEP} {\bfseries 08}
  (2019) 137}, \href{http://arxiv.org/abs/1907.01556}{{\ttfamily
  arXiv:1907.01556 [hep-ph]}}\relax
\mciteBstWouldAddEndPunctfalse
\mciteSetBstMidEndSepPunct{\mcitedefaultmidpunct}
{}{\mcitedefaultseppunct}\relax
\EndOfBibitem
\bibitem{Davier:2019can}
M.~Davier, A.~Hoecker, B.~Malaescu, and Z.~Zhang, ``{A new evaluation of the
  hadronic vacuum polarisation contributions to the muon anomalous magnetic
  moment and to $\mathbf{\boldsymbol\alpha(m_Z^2)}$},''
  \href{http://dx.doi.org/10.1140/epjc/s10052-020-7792-2}{{\em Eur. Phys. J. C}
  {\bfseries 80} no.~3, (2020) 241},
  \href{http://arxiv.org/abs/1908.00921}{{\ttfamily arXiv:1908.00921
  [hep-ph]}}. [Erratum: Eur.Phys.J.C 80, 410 (2020)]\relax
\mciteBstWouldAddEndPunctfalse
\mciteSetBstMidEndSepPunct{\mcitedefaultmidpunct}
{}{\mcitedefaultseppunct}\relax
\EndOfBibitem
\bibitem{Keshavarzi:2019abf}
A.~Keshavarzi, D.~Nomura, and T.~Teubner, ``{$g-2$ of charged leptons, $\alpha
  (M^2_Z)$ , and the hyperfine splitting of muonium},''
  \href{http://dx.doi.org/10.1103/PhysRevD.101.014029}{{\em Phys. Rev. D}
  {\bfseries 101} no.~1, (2020) 014029},
  \href{http://arxiv.org/abs/1911.00367}{{\ttfamily arXiv:1911.00367
  [hep-ph]}}\relax
\mciteBstWouldAddEndPunctfalse
\mciteSetBstMidEndSepPunct{\mcitedefaultmidpunct}
{}{\mcitedefaultseppunct}\relax
\EndOfBibitem
\bibitem{Kurz:2014wya}
A.~Kurz, T.~Liu, P.~Marquard, and M.~Steinhauser, ``{Hadronic contribution to
  the muon anomalous magnetic moment to next-to-next-to-leading order},''
  \href{http://dx.doi.org/10.1016/j.physletb.2014.05.043}{{\em Phys. Lett. B}
  {\bfseries 734} (2014) 144--147},
  \href{http://arxiv.org/abs/1403.6400}{{\ttfamily arXiv:1403.6400
  [hep-ph]}}\relax
\mciteBstWouldAddEndPunctfalse
\mciteSetBstMidEndSepPunct{\mcitedefaultmidpunct}
{}{\mcitedefaultseppunct}\relax
\EndOfBibitem
\bibitem{Melnikov:2003xd}
K.~Melnikov and A.~Vainshtein, ``{Hadronic light-by-light scattering
  contribution to the muon anomalous magnetic moment revisited},''
  \href{http://dx.doi.org/10.1103/PhysRevD.70.113006}{{\em Phys. Rev. D}
  {\bfseries 70} (2004) 113006},
  \href{http://arxiv.org/abs/hep-ph/0312226}{{\ttfamily
  arXiv:hep-ph/0312226}}\relax
\mciteBstWouldAddEndPunctfalse
\mciteSetBstMidEndSepPunct{\mcitedefaultmidpunct}
{}{\mcitedefaultseppunct}\relax
\EndOfBibitem
\bibitem{Masjuan:2017tvw}
P.~Masjuan and P.~Sanchez-Puertas, ``{Pseudoscalar-pole contribution to the
  $(g_{\mu}-2)$: a rational approach},''
  \href{http://dx.doi.org/10.1103/PhysRevD.95.054026}{{\em Phys. Rev. D}
  {\bfseries 95} no.~5, (2017) 054026},
  \href{http://arxiv.org/abs/1701.05829}{{\ttfamily arXiv:1701.05829
  [hep-ph]}}\relax
\mciteBstWouldAddEndPunctfalse
\mciteSetBstMidEndSepPunct{\mcitedefaultmidpunct}
{}{\mcitedefaultseppunct}\relax
\EndOfBibitem
\bibitem{Colangelo:2017fiz}
G.~Colangelo, M.~Hoferichter, M.~Procura, and P.~Stoffer, ``{Dispersion
  relation for hadronic light-by-light scattering: two-pion contributions},''
  \href{http://dx.doi.org/10.1007/JHEP04(2017)161}{{\em JHEP} {\bfseries 04}
  (2017) 161}, \href{http://arxiv.org/abs/1702.07347}{{\ttfamily
  arXiv:1702.07347 [hep-ph]}}\relax
\mciteBstWouldAddEndPunctfalse
\mciteSetBstMidEndSepPunct{\mcitedefaultmidpunct}
{}{\mcitedefaultseppunct}\relax
\EndOfBibitem
\bibitem{Hoferichter:2018kwz}
M.~Hoferichter, B.-L. Hoid, B.~Kubis, S.~Leupold, and S.~P. Schneider,
  ``{Dispersion relation for hadronic light-by-light scattering: pion pole},''
  \href{http://dx.doi.org/10.1007/JHEP10(2018)141}{{\em JHEP} {\bfseries 10}
  (2018) 141}, \href{http://arxiv.org/abs/1808.04823}{{\ttfamily
  arXiv:1808.04823 [hep-ph]}}\relax
\mciteBstWouldAddEndPunctfalse
\mciteSetBstMidEndSepPunct{\mcitedefaultmidpunct}
{}{\mcitedefaultseppunct}\relax
\EndOfBibitem
\bibitem{Gerardin:2019vio}
A.~G\'erardin, H.~B. Meyer, and A.~Nyffeler, ``{Lattice calculation of the pion
  transition form factor with $N_f=2+1$ Wilson quarks},''
  \href{http://dx.doi.org/10.1103/PhysRevD.100.034520}{{\em Phys. Rev. D}
  {\bfseries 100} no.~3, (2019) 034520},
  \href{http://arxiv.org/abs/1903.09471}{{\ttfamily arXiv:1903.09471
  [hep-lat]}}\relax
\mciteBstWouldAddEndPunctfalse
\mciteSetBstMidEndSepPunct{\mcitedefaultmidpunct}
{}{\mcitedefaultseppunct}\relax
\EndOfBibitem
\bibitem{Bijnens:2019ghy}
J.~Bijnens, N.~Hermansson-Truedsson, and A.~Rodr\'\i{}guez-S\'anchez,
  ``{Short-distance constraints for the HLbL contribution to the muon anomalous
  magnetic moment},''
  \href{http://dx.doi.org/10.1016/j.physletb.2019.134994}{{\em Phys. Lett. B}
  {\bfseries 798} (2019) 134994},
  \href{http://arxiv.org/abs/1908.03331}{{\ttfamily arXiv:1908.03331
  [hep-ph]}}\relax
\mciteBstWouldAddEndPunctfalse
\mciteSetBstMidEndSepPunct{\mcitedefaultmidpunct}
{}{\mcitedefaultseppunct}\relax
\EndOfBibitem
\bibitem{Colangelo:2019uex}
G.~Colangelo, F.~Hagelstein, M.~Hoferichter, L.~Laub, and P.~Stoffer,
  ``{Longitudinal short-distance constraints for the hadronic light-by-light
  contribution to $(g-2)_\mu$ with large-$N_c$ Regge models},''
  \href{http://dx.doi.org/10.1007/JHEP03(2020)101}{{\em JHEP} {\bfseries 03}
  (2020) 101}, \href{http://arxiv.org/abs/1910.13432}{{\ttfamily
  arXiv:1910.13432 [hep-ph]}}\relax
\mciteBstWouldAddEndPunctfalse
\mciteSetBstMidEndSepPunct{\mcitedefaultmidpunct}
{}{\mcitedefaultseppunct}\relax
\EndOfBibitem
\bibitem{Colangelo:2014qya}
G.~Colangelo, M.~Hoferichter, A.~Nyffeler, M.~Passera, and P.~Stoffer,
  ``{Remarks on higher-order hadronic corrections to the muon
  g\ensuremath{-}2},''
  \href{http://dx.doi.org/10.1016/j.physletb.2014.06.012}{{\em Phys. Lett. B}
  {\bfseries 735} (2014) 90--91},
  \href{http://arxiv.org/abs/1403.7512}{{\ttfamily arXiv:1403.7512
  [hep-ph]}}\relax
\mciteBstWouldAddEndPunctfalse
\mciteSetBstMidEndSepPunct{\mcitedefaultmidpunct}
{}{\mcitedefaultseppunct}\relax
\EndOfBibitem
\bibitem{Blum:2019ugy}
T.~Blum, N.~Christ, M.~Hayakawa, T.~Izubuchi, L.~Jin, C.~Jung, and C.~Lehner,
  ``{Hadronic Light-by-Light Scattering Contribution to the Muon Anomalous
  Magnetic Moment from Lattice QCD},''
  \href{http://dx.doi.org/10.1103/PhysRevLett.124.132002}{{\em Phys. Rev.
  Lett.} {\bfseries 124} no.~13, (2020) 132002},
  \href{http://arxiv.org/abs/1911.08123}{{\ttfamily arXiv:1911.08123
  [hep-lat]}}\relax
\mciteBstWouldAddEndPunctfalse
\mciteSetBstMidEndSepPunct{\mcitedefaultmidpunct}
{}{\mcitedefaultseppunct}\relax
\EndOfBibitem
\bibitem{Aoyama:2012wk}
T.~Aoyama, M.~Hayakawa, T.~Kinoshita, and M.~Nio, ``{Complete Tenth-Order QED
  Contribution to the Muon g-2},''
  \href{http://dx.doi.org/10.1103/PhysRevLett.109.111808}{{\em Phys. Rev.
  Lett.} {\bfseries 109} (2012) 111808},
  \href{http://arxiv.org/abs/1205.5370}{{\ttfamily arXiv:1205.5370
  [hep-ph]}}\relax
\mciteBstWouldAddEndPunctfalse
\mciteSetBstMidEndSepPunct{\mcitedefaultmidpunct}
{}{\mcitedefaultseppunct}\relax
\EndOfBibitem
\bibitem{atoms7010028}
T.~Aoyama, T.~Kinoshita, and M.~Nio, ``Theory of the anomalous magnetic moment
  of the electron,'' \href{http://dx.doi.org/10.3390/atoms7010028}{{\em Atoms}
  {\bfseries 7} no.~1, (2019) }.
  \url{https://www.mdpi.com/2218-2004/7/1/28}\relax
\mciteBstWouldAddEndPunctfalse
\mciteSetBstMidEndSepPunct{\mcitedefaultmidpunct}
{}{\mcitedefaultseppunct}\relax
\EndOfBibitem
\bibitem{Czarnecki:2002nt}
A.~Czarnecki, W.~J. Marciano, and A.~Vainshtein, ``{Refinements in electroweak
  contributions to the muon anomalous magnetic moment},''
  \href{http://dx.doi.org/10.1103/PhysRevD.67.073006}{{\em Phys. Rev. D}
  {\bfseries 67} (2003) 073006},
  \href{http://arxiv.org/abs/hep-ph/0212229}{{\ttfamily arXiv:hep-ph/0212229}}.
  [Erratum: Phys.Rev.D 73, 119901 (2006)]\relax
\mciteBstWouldAddEndPunctfalse
\mciteSetBstMidEndSepPunct{\mcitedefaultmidpunct}
{}{\mcitedefaultseppunct}\relax
\EndOfBibitem
\bibitem{Gnendiger:2013pva}
C.~Gnendiger, D.~St\"ockinger, and H.~St\"ockinger-Kim, ``{The electroweak
  contributions to $(g-2)_\mu$ after the Higgs boson mass measurement},''
  \href{http://dx.doi.org/10.1103/PhysRevD.88.053005}{{\em Phys. Rev. D}
  {\bfseries 88} (2013) 053005},
  \href{http://arxiv.org/abs/1306.5546}{{\ttfamily arXiv:1306.5546
  [hep-ph]}}\relax
\mciteBstWouldAddEndPunctfalse
\mciteSetBstMidEndSepPunct{\mcitedefaultmidpunct}
{}{\mcitedefaultseppunct}\relax
\EndOfBibitem
\bibitem{Sirunyan:2021khd}
{\bfseries CMS} Collaboration, A.~M. Sirunyan {\em et~al.}, ``{Search for
  resonant and nonresonant new phenomena in high-mass dilepton final states at
  $\sqrt{s} = $ 13 TeV},'' \href{http://arxiv.org/abs/2103.02708}{{\ttfamily
  arXiv:2103.02708 [hep-ex]}}\relax
\mciteBstWouldAddEndPunctfalse
\mciteSetBstMidEndSepPunct{\mcitedefaultmidpunct}
{}{\mcitedefaultseppunct}\relax
\EndOfBibitem
\bibitem{Aad:2019fac}
{\bfseries ATLAS} Collaboration, G.~Aad {\em et~al.}, ``{Search for high-mass
  dilepton resonances using 139 fb$^{-1}$ of $pp$ collision data collected at
  $\sqrt{s}=$13 TeV with the ATLAS detector},''
  \href{http://dx.doi.org/10.1016/j.physletb.2019.07.016}{{\em Phys. Lett. B}
  {\bfseries 796} (2019) 68--87},
  \href{http://arxiv.org/abs/1903.06248}{{\ttfamily arXiv:1903.06248
  [hep-ex]}}\relax
\mciteBstWouldAddEndPunctfalse
\mciteSetBstMidEndSepPunct{\mcitedefaultmidpunct}
{}{\mcitedefaultseppunct}\relax
\EndOfBibitem
\bibitem{Aad:2020iuy}
{\bfseries ATLAS} Collaboration, G.~Aad {\em et~al.}, ``{Search for pairs of
  scalar leptoquarks decaying into quarks and electrons or muons in $ \sqrt{s}
  $ = 13 TeV $pp$ collisions with the ATLAS detector},''
  \href{http://dx.doi.org/10.1007/JHEP10(2020)112}{{\em JHEP} {\bfseries 10}
  (2020) 112}, \href{http://arxiv.org/abs/2006.05872}{{\ttfamily
  arXiv:2006.05872 [hep-ex]}}\relax
\mciteBstWouldAddEndPunctfalse
\mciteSetBstMidEndSepPunct{\mcitedefaultmidpunct}
{}{\mcitedefaultseppunct}\relax
\EndOfBibitem
\end{mcitethebibliography}
\end{document}